\documentclass[twocolumn]{aastex631}

\usepackage{ifthen}
\usepackage{afterpage}

\newcommand{\comment}[1]{}

\newcommand{\change}[1]{#1}

\usepackage{amsmath}
\usepackage{natbib}
\usepackage{appendix}
\usepackage{gensymb}

\providecommand{\adsurl}[1]{\href{#1}{ADS}}

\DeclareSymbolFont{UPM}{U}{eur}{m}{n}
\DeclareMathSymbol{\umu}{0}{UPM}{"16}
\let\oldumu=\umu
\renewcommand\umu{\ifmmode\oldumu\else$\oldumu$\fi}
\newcommand\micro{\umu}

\newcommand\microns{\micro m}

\definecolor{color1}{HTML}{000000} 
\definecolor{color2}{HTML}{313695} 
\definecolor{color3}{HTML}{FD8208} 
\definecolor{color4}{HTML}{A50026} 

\comment{\hfill\textbf{DRAFT of {\today} \now}.}

\shorttitle{ThERESA: 3D Eclipse Mapping}
\shortauthors{Challener \& Rauscher}

\begin{document}

\title{ThERESA: Three-Dimensional Eclipse Mapping with Application to Synthetic JWST Data}
\author[0000-0002-8211-6538]{Ryan C. Challener}
\affiliation{Department of Astronomy, University of Michigan, 1085
  S. University Ave., Ann Arbor, MI 48109, USA}

\author[0000-0003-3963-9672]{Emily Rauscher}
\affiliation{Department of Astronomy, University of Michigan, 1085
  S. University Ave., Ann Arbor, MI 48109, USA}


\begin{abstract}

  Spectroscopic eclipse observations, like those possible with the
  \textit{James Webb Space Telescope}, should enable 3D mapping of
  exoplanet daysides. However, fully-flexible 3D planet models are
  overly complex for the data and computationally infeasible for
  data-fitting purposes. Here, we present ThERESA, a method to
  retrieve the 3D thermal structure of an exoplanet from eclipse
  observations by first retrieving 2D thermal maps at each wavelength
  and then placing them vertically in the atmosphere. This approach
  allows the 3D model to include complex thermal structures with a
  manageable number of parameters, hastening fit convergence and
  limiting overfitting. An analysis runs in a matter of days. We
  enforce consistency of the 3D model by comparing vertical placement
  of the 2D maps with their corresponding contribution functions.  To
  test this approach, we generated a synthetic JWST NIRISS-like
  observation of a single hot-Jupiter eclipse using a global
  circulation model of WASP-76b and retrieved its 3D thermal
  structure. We find that a model which places the 2D maps at
  different depths depending on latitude and longitude is preferred
  over a model with a single pressure for each 2D map, indicating that
  ThERESA is able to retrieve 3D atmospheric structure from JWST
  observations. We successfully recover the temperatures of the
  planet's dayside, the eastward shift of its hotspot, and the thermal
  inversion. ThERESA is open-source and publicly available as a tool
  for the community.

\end{abstract}


\section{INTRODUCTION}
\label{sec:introduction}

Exoplanet atmospheric retrieval is a method of \change{inferring} the
temperature and composition of extrasolar planets \change{based on the 
observed spectrum and photometry}. Retrieval can be
done in two ways: by modeling planetary emission, measured through
direct observation or monitoring flux loss during eclipse, and by
modeling the transmission of stellar light through the planet's
atmosphere while the planet transits its host star
\citep[e.g.,][]{DemingSeager2017JGREreview}. Historically, due to the
challenges of observing planets in the presence of much larger,
brighter stars, exoplanet retrievals have been limited to measurement
of bulk properties, using a single temperature-pressure profile and
set of molecular abundance profiles for the entire planet
\citep[e.g.,][]{KreidbergEtal2015WASP12b, HardyEtal2017apjHATP13b,
  GarhartEtal2018aapQatar1b}. Analyses like these can be biased, with
the retrieved properties possibly not representative of any single
location on the planet \citep{FengEtal2016apj1DBias, LineParmentier2016apjPatchyClouds,
  BlecicEtal2017apj1DBias, LacyBurrows2020apj3DBias, MacDonadEtal2020apjl1DBias, TaylorEtal2020mnras3DBias}.

Eclipse mapping is a technique for converting exoplanet light curves
to brightness maps. During eclipse ingress and egress, the planet's
host star blocks and uncovers different slices of the planet in time.
Brightness variations across the planet result in changes in the
morphology of the eclipse light curve
\citep{WilliamsEtal2006apjEclipseMapping, RauscherEtal2007apjEclipseMapping, 
CowanFujii2018haexMappingReview}. HD 189733 b is the only
planet successfully mapped from eclipse observations, by stacking many
8.0 \microns\ light curves \citep{DeWitEtal2012aaHD189Map,
  MajeauEtal2012apjlHD189Map}. However, with the advent of the
\textit{James Webb Space Telescope} (JWST) in the near future, we
expect observations of many more planets to be of sufficient quality
for eclipse mapping \citep{SchlawinEtal2018ajJWSTmapping}.

2D eclipse mapping, where a single light curve is inverted to a
spatial brightness map, is a complex process. Depending on assumptions
about the planet's structure, retrieved maps can be strongly
correlated with orbital and system parameters
\citep{DeWitEtal2012aaHD189Map}. \cite{RauscherEtal2018ajMap}
presented a mapping technique using an orthogonal basis of light
curves, reducing parameter correlations and extracting the maximum
information possible.

In principle, spectroscopic eclipse observations, like those possible
with JWST, should allow 3D eclipse mapping of exoplanets. Every
wavelength probes different pressures in the planet's
atmosphere. These ranges depend on the wavelength-dependent opacity of
the atmosphere, which in turn depends on the absorption, emission, and
scattering properties of the atmosphere's constituents. Thus, eclipse
observations are sensitive to both temperature and composition as
functions of latitude and longitude. In practice, however, 3D eclipse
mapping is complex. Each thermal map computed from a spectral light
curve corresponds to a range or ranges of pressures, which can vary
significantly across the planet
\citep{Dobbs-DixonCown2017apjContributions}. Unlike 2D mapping at a
single wavelength, 3D mapping requires computationally-expensive
radiative transfer calculations, which can be prohibitive when
exploring model parameter space. Also, the 3D model parameter space is
extensive, so one must make simplifying considerations based on the
quality of the data.

\cite{MansfieldEtal2020mnrasEigenspectraMapping} introduced a method
to use clustering algorithms with a set of multi-wavelength maps to
divide the planet into several regions with similar spectra. While not 
a fully-3D model, this method shows promise for distinguishing spatial 
regions with distinct thermal profiles or chemical compositions, determined
from atmospheric retrieval of the individual regions.

In this work we build upon \cite{RauscherEtal2018ajMap}, combining
eclipse mapping techniques with 1D radiative transfer to present a
mapping method that captures the full 3D nature of exoplanet
atmospheres while being maximally informative, with constraints to
enforce physical plausibility and considerations that improve runtime.
In Section \ref{sec:methods} we describe our 2D and 3D mapping
approaches, in Section \ref{sec:w76b} we apply our methods to
synthetic observations, in Section \ref{sec:disc} we compare our
retrieved 2D and 3D maps with the input planet model, and in Section
\ref{sec:conc} we summarize our conclusions.

\section{METHODS}
\label{sec:methods}

\begin{figure*}
    \includegraphics[width=7in]{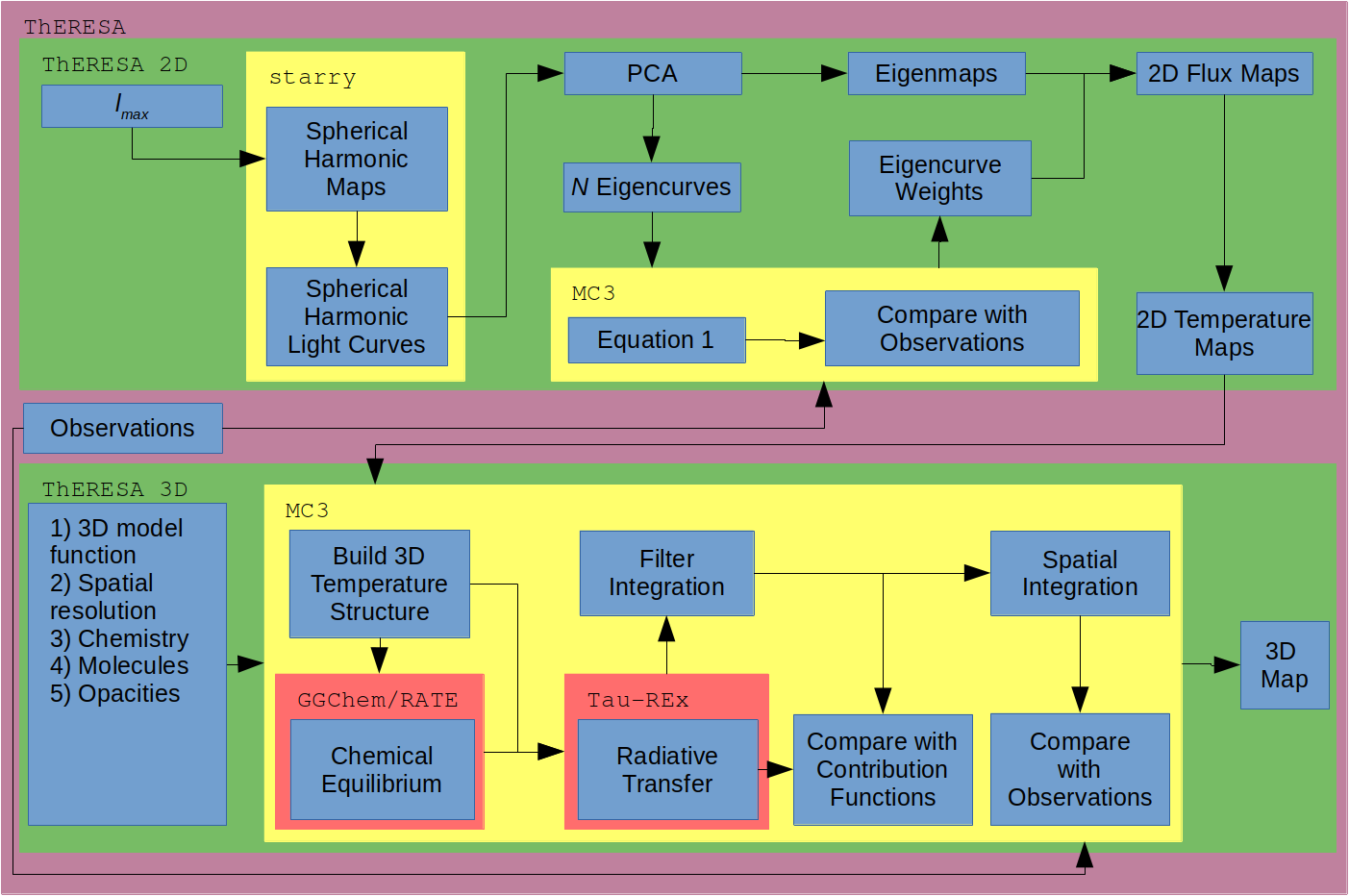}
    \caption{ \label{fig:diagram} The code structure of ThERESA. The
    blue boxes show inputs or stages of the analysis, with arrows
    showing the order of execution. The background
    colors show which code(s) perform each stage. The steps within
    the MC3 boxes are run thousands of times by MC3, but the
    calculations are handled by ThERESA (unless otherwise noted).}
\end{figure*}

Here we present the Three-dimensional Exoplanet Retrieval from Eclipse
Spectroscopy of Atmospheres code
(ThERESA\footnote{https://github.com/rychallener/ThERESA}). ThERESA
combines the methods of \cite{RauscherEtal2018ajMap} with
thermochemical equilibrium calculations, radiative transfer, and
planet integration to simultaneously fit spectroscopic eclipse light
curves, retrieving the three-dimensional thermal structure of
exoplanets. The code operates in two modes -- 2D and 3D -- with the
former as a pre-requisite for the latter. \change{The code structure
  is shown in Figure \ref{fig:diagram}, with further description in
  the following sections. The version of ThERESA used for this
  analysis can be found at
  \url{https://doi.org/10.5281/zenodo.5773215}.}

\subsection{2D Mapping}

ThERESA's 2D mapping follows the methods of
\cite{RauscherEtal2018ajMap}. First, we calculate a basis of light
curves, at the supplied observation times, from positive and negative
spherical-harmonic maps $Y^l_m$, up to a user-supplied complexity
$l_{max}$ using the \texttt{starry} package
\citep{LugerEtal2019ajStarry}. These light curves are then run through
a principle component analysis (PCA) to determine a new basis set of
orthogonal light curves, ordered by total power, which are used in a
linear combination to individually model the spectroscopic light
curves.

\begin{figure*}
  \includegraphics[width=7in]{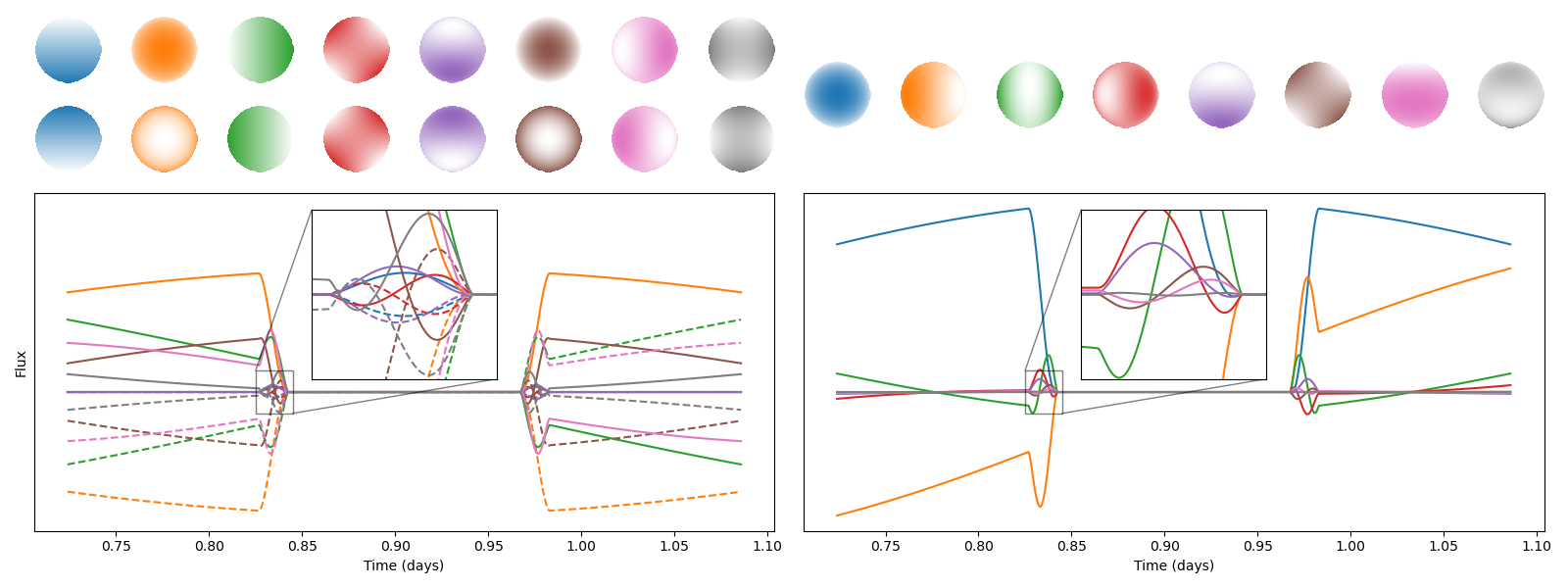}
  \caption{\label{fig:eigencurves} \textbf{Top Left:} The positive and
    negative spherical harmonic maps with $l \leq 2$, omitting the
    uniform map $Y^0_0$. \textbf{Bottom Left:} The secondary-eclipse 
    light curves from
    integrating the spherical harmonic maps in the top left. Light
    curve colors match corresponding maps. Dashed lines correspond to
    negated spherical harmonic maps. \textbf{Bottom Right:} The
    eigencurves with the eight largest eigenvalues resulting from truncated
    singular-value decomposition principle component analysis applied to
    the spherical-harmonic curves in the bottom left. 
    Eigenmaps are ordered, left to right, from most to least 
    information content. 
    \textbf{Top Right:} The eigenmaps corresponding to 
    each eigencurve in the bottom right, with matching colors.}
\end{figure*}

In typical PCA, one subtracts the mean from each observation
(spherical-harmonic light curve), computes the covariance matrix of
the mean-subtracted set of observations, then calculates the
eigenvalues and eigenvectors of this covariance matrix. The new set of
observations (``eigencurves'') is the dot product of the eigenvectors
and the mean-subtracted light curves, and the eigenvalues are the
contributions from each spherical harmonic map to generate each
eigencurve. However, the mean-subtraction causes the initial light
curve basis set to have non-zero values during eclipse, a physical
impossibility that propagates forward to the new basis set of
orthogonal light curves. Integrating a map (an ``eigenmap'') created
from these eigenvalues will not generate a light curve that matches
the eigencurves. Therefore, we use truncated singular-value
decomposition (TSVD), provided by the \texttt{scikit-learn} package
\citep{PedregosaEtal2011jmlrScikitLearn}. TSVD does not do
mean-subtraction so the resulting eigencurves have zero flux during
eclipse, as expected, which is an improvement over 
\cite{RauscherEtal2018ajMap}. Figure \ref{fig:eigencurves} shows an example of
the transformation from spherical-harmonic maps and light curves to
eigenmaps and eigencurves.

We fit each star-normalized spectroscopic light curve, individually, as a
linear combination of $N$ eigencurves, the uniform-map light curve
$Y_0^0$, and a constant offset $s_{corr}$ to account for any
stellar flux normalization errors. Each wavelength has its own
set of fit values, and potentially its own set of eigencurves, since
$N$ and $l_{max}$ can be different for each light curve. Functionally, 
the model is

\begin{equation}
  F_{sys}(t) = c_0 Y_0^0 + \sum_i^N c_iE_i + s_{corr},
\end{equation}

\noindent
where $F_{sys}$ is the system flux, $c_i$ are the light-curve weights,
and $E_i$ are the eigencurves. The light-curve weights and $s_{corr}$
are the free parameters of the model. Although the synthetic data in
this work should have $s_{corr} = 0$, we still fit to this parameter
to better represent an analysis of real data. We run a $\chi^2$
minimization to determine the best fit and then run Markov-chain Monte
Carlo (MCMC), through MC3 \citep{CubillosEtal2017ajRedNoise}, to fully
explore the parameter space.

By construction, the eigenmaps and eigencurves are deviations from the
uniform map and its corresponding light curve, respectively, so
negative values are possible. If the parameter space is left
unconstrained, some regions of the planet may be best fit with a
negative flux. To avoid this non-physical scenario, we check for
negative fluxes across the visible cells of the planet, based on the
times of observation, and penalize the fit. The penalty scales with
the magnitude of the negative flux, and we ensure that any negative
fluxes result in a worse fit than any planet with all positive fluxes,
which effectively guides the fit away from non-physical planets. While
the eigenmaps seemingly provide information about the non-visible
cells of the planet, these constraints are simply a consequence of the
continuity of spherical harmonic maps. Given that we have no real
information on those portions of the planet, we do not insist that
they have positive fluxes.

We then use the best-fitting parameters with the matching eigenmaps to
construct a thermal flux map for the light curve observed at each wavelength
(Equation 4 of \citealp{RauscherEtal2018ajMap}):

\begin{equation}
  Z_p(\theta, \phi) = c_0 Y_0^0(\theta, \phi) + \sum_i^N c_i Z_i(\theta, \phi),
\end{equation}

\noindent
where $Z_p$ is the thermal flux map, $\theta$ is latitude, $\phi$ is
longitude, and $Z_i$ are the eigenmaps. These flux maps are converted
to temperature maps\footnote{These are brightness temperatures, but
we later treat them as physical temperatures at the pressures of each
wavelength's contribution function.} using Equation 8 of
\cite{RauscherEtal2018ajMap}:

\begin{equation}
  \label{eqn:fmaptotmap}
  T_p(\theta, \phi) = (hc/\lambda k) / \textrm{ln} \left[ 1 + \left(\frac{R_p}{R_s}\right)^2 \frac{\exp[hc/\lambda k T_s] - 1}{\pi Z_p(\theta, \phi)(1 + s_{corr})}\right],
\end{equation}

\noindent
where $\lambda$ is the band-averaged wavelength of the filter used to
observe the corresponding light curve, $R_p$ is the radius of the
planet, $R_s$ is the radius of the star, and $T_s$ is the stellar
temperature.

When performing 2D mapping, the primary user decisions are choosing
$l_{max}$, the maximum order of the spherical harmonic light curves,
and $N$, the number of eigencurves to include in the fit. Larger
$l_{max}$ (up to a limit) and $N$ result in better fits, as more
complex thermal structures become possible. However, these complex
planets are often not justified by the quality of the data. To choose
$l_{max}$ and $N$ we use the Bayesian Information Criterion (BIC,
\citealp{Raftery1995BIC}), defined as

\begin{equation}
  \label{eqn:bic}
  \textrm{BIC} = \chi^2 + k \ln n_{data},
\end{equation}

\noindent
where $\chi^2$ is the traditional goodness-of-fit metric, $k$ is the
number of free parameters in the model ($N+2$ per light curve,
assuming a uniform map term $c_0$ and $s_{\textrm{corr}}$), and
$n_{data}$ is the number of data points being fit. Thus, the BIC
penalizes fits which are overly complex. We choose the 2D fit which
results in the lowest BIC as the best fit.

\subsubsection{Application to Spitzer HD 189733b Observations}

To test our implementation of the methods in
\cite{RauscherEtal2018ajMap}, and the effects of TSVD PCA, we
performed the same analysis of \textit{Spitzer} phase curve and
eclipse observations of HD 189733b. The data are the same as those
used by \cite{MajeauEtal2012apjlHD189Map}, which include eclipse
observations \citep{AgolEtal2010apjHD189} and approximately a quarter
of a phase curve (\citealp{KnutsonEtal2007natHD189phasecurve},
re-reduced by \citealp{AgolEtal2010apjHD189}).

\begin{figure*}
  \includegraphics[width=7in]{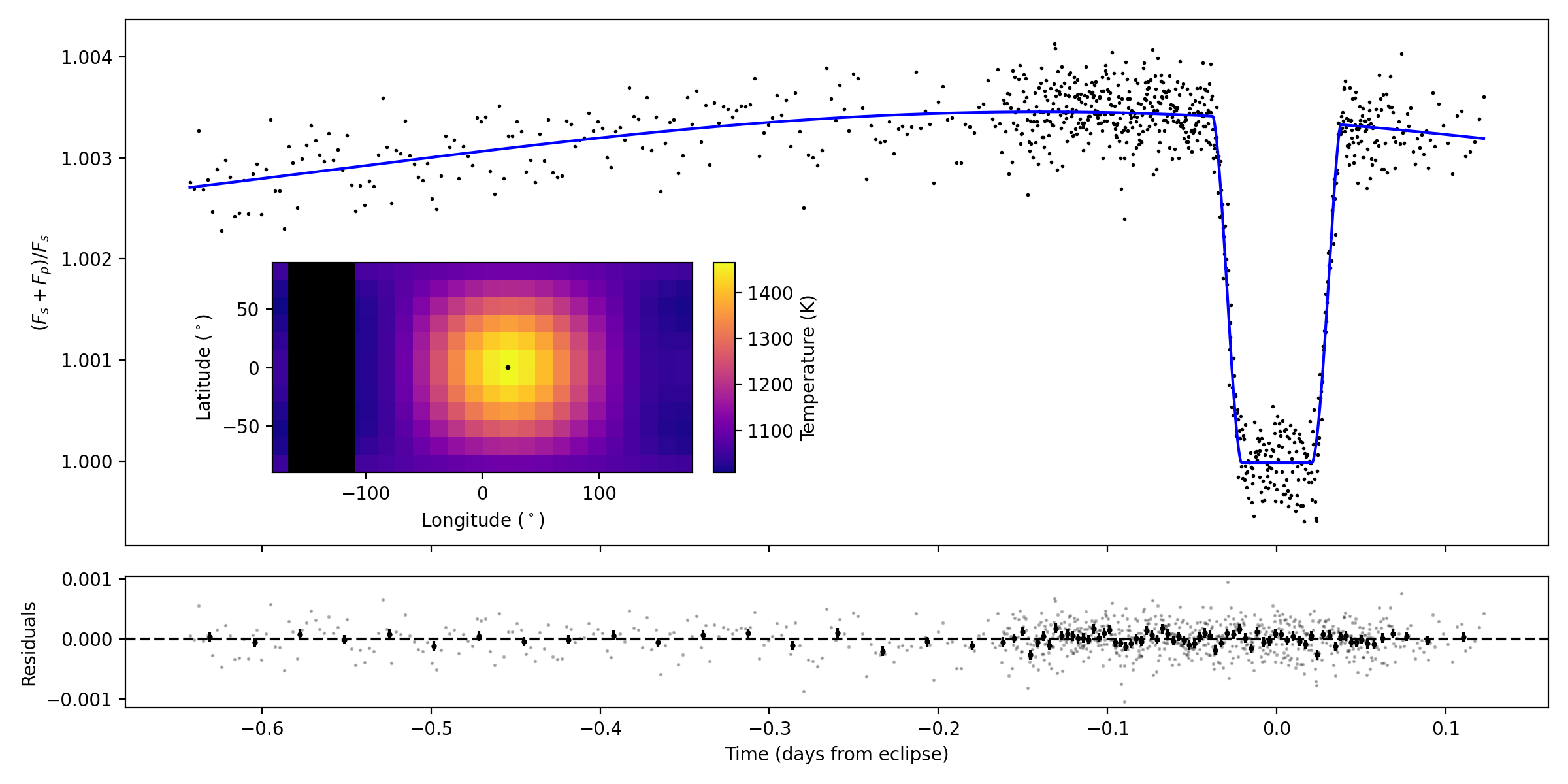}
  \caption{\label{fig:hd189} Results of a 2D fit to the HD 189733b
    \textit{Spitzer} observations. \textbf{Top:} The light-curve data
    and best-fitting model. The inset shows the best-fitting thermal
    map. A black box covers the longitudes not visible during the
    observations. The hotspot location is marked with a black
    dot. \textbf{Bottom:} The model residuals and uncertainties. Data
    are binned for visual clarity, to 10 data points per bin.}
\end{figure*}

Like \cite{RauscherEtal2018ajMap}, we tested a range of values for
$l_{\rm max}$ and $N$, using the BIC to choose the model with the
highest complexity justified by the data. Table \ref{tbl:hd189}
compares our goodness-of-fit statistics with
\cite{RauscherEtal2018ajMap}. We also determine that the optimal fit
uses $l_{\rm max} = 2$ and $N = 2$ (see figure \ref{fig:hd189}), and
our preference for $N = 2$ over $N = 3$ is stronger. As expected,
models with higher $N$ result in lower $\chi^2$.  For low $N$, we
achieve better $\chi^2$ and BIC values than
\cite{RauscherEtal2018ajMap}, although the difference is slight at
$N=2$. At higher $N$, differences between the results are
statistically negligible. These differences are likely due to the PCA
methods used and any differences in the spherical-harmonic light-curve
calculation packages used, since \cite{RauscherEtal2018ajMap} employed
SPIDERMAN \citep{LoudenKreidberg2018mnrasSPIDERMAN}.

\begin{deluxetable}{cccccc}
  \tablecolumns{6}
  \tablecaption{HD 189733b Goodness-of-fit Statistics \label{tbl:hd189}}
  \tablehead{
    & & \multicolumn{2}{c}{This work} & \multicolumn{2}{c}{\cite{RauscherEtal2018ajMap}} \\
    \colhead{$l_{\rm max}$} & \colhead{$N$} & \colhead{$\chi^2$} & \colhead{BIC} & \colhead{$\chi^2$} & \colhead{BIC}
  }
    \startdata
    2 & 1 & 952.7 & 973.1 & 1054  & 1074\\
    2 & 2 & 938.2 & 965.3 & 942.4 & 969.6 \\
    2 & 3 & 938.0 & 971.9 & 938.8 & 972.7 \\
    2 & 4 & 937.5 & 978.2 & 936.4 & 977.1 \\
    2 & 5 & 937.4 & 984.8 & 936.4 & 984.0 \\
    \enddata
\end{deluxetable}

Since our eigencurves differ from those used by
\cite{RauscherEtal2018ajMap}, there is no meaningful comparison
between the best-fitting $c_i$ parameters. However, the stellar
correction $s_{\rm corr}$ has the same function in both works.
Notably, we find $s_{\rm corr} = -14 \pm 20$ ppm, consistent with
zero, whereas \cite{RauscherEtal2018ajMap} find a stellar correction
of $452^{+39}_{-40}$ ppm. It is likely that our use of TSVD PCA to
enforce zero flux during eigencurve eclipse means the stellar
correction term is only correcting for normalization errors, and not
also ensuring zero planet flux during eclipse.

To determine the location of the planet's hotspot, we use
\texttt{starry} to find the location of maximum brightness of the
best-fit 2D thermal map. Then, to calculate an uncertainty, we repeat
this process on 10,000 maps sampled from the MCMC posterior
distribution, taking the standard deviation of these locations to be
the uncertainty. For HD 189733b, we find a \change{hotspot} longitude of $21.8
\pm 1.5^\circ$ to the east of the substellar point, which very closely 
matches \change{the $21.6 \pm 1.6^\circ$ and $21.8 \pm 1.5^\circ$} found by
\cite{RauscherEtal2018ajMap} and \cite{MajeauEtal2012apjlHD189Map}\change{, respectively}.
Thus, we are confident that we have accurately implemented 2D mapping.

\subsection{3D Mapping}
\label{sec:method3d}

Atmospheric opacity varies with wavelength, so spectrosopic eclipse
observations probe multiple depths of the planet, in principle
allowing for three-dimensional atmospheric retrieval. However, the
relationship between wavelength and pressure is very complex. Each
wavelength probes a range or ranges of pressures, and those pressures
change with location on the planet
\citep{Dobbs-DixonCown2017apjContributions}. Here, we combine our 2D
maps vertically within an atmosphere, run radiative transfer,
integrate over the planet, and compare to all spectroscopic light
curves simultaneously with MCMC to retrieve the 3D thermal structure
of the planet and parameter uncertainties. We assume that the planet's
atmosphere is static, so that observation time only affects viewing
geometry.

The complexity of the radiative transfer calculation and the resulting
computation cost forces us to discard the arbitrary resolution of the
2D maps in favor of a gridded planet. The grid size is up to the user
(see Section \ref{sec:gridsize}, but smaller grid cells quickly increase 
mapping runtime.

The 3D planet model is the relationship between 2D temperature maps
and pressures. ThERESA offers several such models:

\begin{enumerate}
\item ``isobaric'' -- The simplest model, where each 2D map is placed
  at a single pressure across the entire planet. This model has one
  free parameter, a pressure (in log space), for each 2D map.

\item ``sinusoidal'' -- A model that allows the pressures of each 2D
  map to vary as a sinusoid with respect to latitude and
  longitude. This model has four free parameters for each 2D map: a
  base pressure, a latitudinal pressure variation amplitude, a
  longitudinal pressure variation amplitude, and a longitudinal shift,
  since hot Jupiters often exhibit hotspot offsets. Functionally,
  the model is:

  \begin{equation}
    \label{eqn:sinusoidal}
    \log{p(\theta, \phi)} = a_1 + a_2 \cos\theta + a_3 \cos(\phi-a_4)
  \end{equation}

  where $a_i$ are free parameters of the model. For simplicity, this
  model does not allow for latitudinal asymmetry, although if the
  2D temperature maps are asymmetric in that manner (detectable in
  inclined orbits), we might expect a similar asymmetry in these pressure
  maps.

\item ``quadratic'' -- A second-degree polynomial model, including
  cross terms, for a total of six free parameters. The functional
  form is:

  \begin{equation}
    \label{eqn:quadratic}
    \begin{split}
      \log{p(\theta, \phi)} = a_1 + a_2\theta^2 + a_3\theta + \\
      a_4\phi^2 + a_5\phi + a_6\theta\phi
    \end{split}
  \end{equation}

  Unlike the isobaric and sinusoidal models, the quadratic model is
  not continuous opposite of the substellar point, where latitude
  rolls over from 180$^\circ$ to -180$^\circ$. This is not an issue
  for eclipse observations, like those presented in this work, but may
  be problematic if all phases of the planet are visible in an
  observation. A similar problem occurs at the poles, although
  these regions contribute very little to the observed flux.

\item ``flexible'' -- A maximally-flexible model that allows each 2D
  map to be placed at an arbitrary pressure for each grid cell. The
  number of free parameters is the number of visible grid cells
  multiplied by the number of 2D maps.
\end{enumerate}

\noindent
These models represent a range of complexities, and the appropriate
function can be chosen using the BIC (Equation \ref{eqn:bic}),
assuming that comparisons are made between fits to the same data.

Once the temperature maps are placed vertically, ThERESA interpolates
each grid cell's 1D thermal profile in logarithmic pressure space. The
interpolation can be linear, a quadratic spline, or a cubic spline. For
pressures above and below the placed temperature maps, temperatures
can be extrapolated, set to isothermal with temperatures equal to the
closest temperature map, or parameterized\change{, as $T_{top}$ and 
$T_{bot}$}. When parameterized, the top
and/or bottom of the atmosphere are given single temperatures across
the entire planet and the pressures between are interpolated according
to the chosen method. If any negative temperatures are found in the
atmosphere grid cells that are visible during the observation, we
discard the fit to avoid these non-physical models.

In principle, atmospheric constituents could also be fitted
parameters. If the data are of sufficient quality, molecular abundance
and temeprature variations across the planet would create observable
emission variability depending on the opacity of the atoms and
molecules. However, without real 3D-capable data to test against, we
assume the planet's chemistry is in thermochemical equilibrium, with
solar atomic abundances. ThERESA offers two ways to calculate
thermochemical equilibrium:
\texttt{rate} \footnote{https://github.com/pcubillos/rate}
\citep{CubillosEtal2019apjRATE} and
GGchem \footnote{https://github.com/pw31/GGchem}
\citep{WoitkeEtal2018aandaGGchem}. With \texttt{rate}, ThERESA
calculates abundances on-the-fly as needed. When using GGchem, the
user must supply a pre-computed grid of abundances over an appropriate
range of temperatures and at the same pressures of the atmosphere used
in the the 3D fit, which ThERESA interpolates as necessary. ThERESA
was designed with flexibility in mind, so other abundance
prescriptions can be inserted.

Next, ThERESA runs a radiative transfer calculation on each grid cell
to compute planetary emission. We use TauREx
3 \footnote{https://github.com/ucl-exoplanets/TauREx3\_public}
\citep{Al-RefaieEtal2019arxivTauRExIII} to run these forward
models. Like the atmospheric abundances, the forward model can be
replaced with any similar function. We only run radiative transfer
(and thermochemical equilibrium) on visible cells to reduce
computation time. Spectra are computed at a higher resolution then
integrated over the observation filters (``tophat'' filters for
spectral bins). For computational feasibility, we use the ExoTransmit
\citep{KemptonEtal2017paspExoTransmit} molecular opacities
\citep{FreedmanEtal2008apjsExoTransmitOpac1,
  FreedmanEtal2014apjsExoTransmitOpac2,
  LupuEtal2014apjExoTransmitOpac3}, \change{which have a resolution of
  10$^3$,} but we note that using
high-resolution line lists such as HITRAN/HITEMP
\citep{RothmanEtal2010jqsrtHITEMP, GordonEtal2017jqsrtHITRAN} or
ExoMol \citep{TennysonEtal2020jqsrtExoMol} would be preferred for
accuracy. In this work we use wide spectral bands, which minimizes the
effect of using low-resolution opacities.

Finally, we integrate over the planet at the observation geometry of
each time in the light curves. The visibility $V$ of each cell,
computed prior to modeling to save computation time, is the integral
of the area and incident angles combined with the blocking effect of
the star, given by

\begin{equation}
  \label{eqn:vis}
  V(\theta, \phi, t) = \begin{cases}
    0 & \text{if}\, d < R_s, \\
    \int_{\theta_i}^{\theta_f} \int_{\phi_i}^{\phi_f} R_p^2 \cos^2 \theta' \cos \phi' d\theta' d\phi' & \text{otherwise}, \\
    \end{cases}
  \end{equation}

\noindent
where $d$ is the projected distance between the center of the visible portion
of a grid cell and the center of the star, defined as

\begin{equation}
  \begin{split}
    d = [(x_p + R_p \cos\theta\sin\phi - x_s)^2 + \\
      (y_p + R_p \sin\theta - y_s)^2]^{1/2}.
    \end{split}
\end{equation}

\noindent
$x_p$ is the $x$ position of the planet, $x_s$ is the $x$ position of
the star, $y_p$ is the $y$ position of the planet, and $y_s$ is the
$y$ position of the star, calculated by \texttt{starry} for each
observation time. \change{These positions include the effect of 
inclination.} We multiply the planetary emission grid by $V$ and
sum at each observation time to calculate the planetary light curve.
Like the 2D fit, we repeat this process in MCMC to fully explore
parameter space. Due to model complexity, traditional least-squares
fitting does not converge to a good fit, so we rely on MCMC for both
model optimization and parameter space exploration.

\subsubsection{Enforcing Contribution Function Consistency}
\label{sec:cffit}

Given absolute freedom to explore the parameter space, the 3D model
will often find regions of parameter space which, while resulting in a
``good'' fit to the spectroscopic light curves, are physically
implausible upon further inspection. For instance, the model may bury
some of the 2D maps deep in the atmosphere where they have no effect
on the planetary emission, or maps may be hidden vertically between
other maps, where the combination of linear interpolation and discrete
pressure layers causes them to not affect the thermal structure.  To
avoid scenarios like these, ThERESA has an option where maps are
required to remain close (in pressure) to their corresponding
contribution functions.

Contribution functions show how much each layer of the atmosphere is
contributing to the flux emitted at the top of the atmosphere as a
function of wavelength \citep[e.g.][]{KnutsonEtal2009apjHD189pc}. When
integrated over a filter's transmission curve, they show which layers
contribute to the emission in that filter. That is, the contribution
function for a given filter shows which pressures of the atmosphere
are probed by an observation with that filter. Therefore, a 2D thermal
map should be placed at a pressure near the peak contribution, or at
least near the integrated average pressure, weighted by the
contribution function for the corresponding filter.

To enforce this condition, we apply a penalty on the 3D model's
$\chi^2$. For each visible grid cell on the planet, we calculate the
contribution function for each filter. We then treat this contribution
function, in log pressure space, as a probability density function and
compute the 68.3\% credible region. Then, we take half the width of
this region to be an approximate 1$\sigma$ width of the contribution
function, and the midpoint of this region to be the approximate
``correct'' location for the map. From this width and location, we
compute a $\chi^2$ for each contribution function and for each grid
cell, and add that to the light-curve $\chi^2$. We caution that this
effectively adds a number of data points equal to the number of
visible grid cells, which means BIC comparisons between models that
use contribution function fitting and have different grid sizes are
invalid.

\section{Application to Synthetic WASP-76b Observations}
\label{sec:w76b}

\subsection{Light-curve Generation}

\begin{figure}
  \includegraphics[width=3.4in]{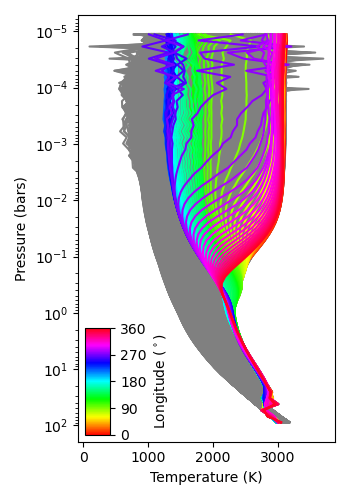}
  \caption{The temperature grid used in light-curve generation. The
    colored profiles show temperatures at various longitudes along the
    equator. The substellar point is $0^\circ$. All other thermal 
    profiles are plotted in gray.\label{fig:gcm}}
\end{figure}

In lieu of real observations, and as a ground-truth test case, we generated synthetic
spectroscopic eclipse light curves of WASP-76b, an ultra-hot Jupiter
with a strong eclipse signal. For the thermal structure, we adopt the
results of a double-grey general circulation model
\citep[GCM,][]{RauscherMenou2012apjGCM,
  RomanRauscher2017apjGCMupdate} from \cite{BeltzEtal2021arxivWASP76bMagField}, with
  no magnetic field. This thermal structure
is shown in Figure \ref{fig:gcm}. The GCM output has 48 latitudes, 96
longitudes, and 65 pressure layers ranging from 100 -- 0.00001 bar in
log space. \change{The double-grey absorption coefficients used in the GCM were chosen to roughly reproduce the average temperature profile for this planet using observational constraints from \cite{FuEtal2021ajWASP76b}, which
includes a thermal inversion, likely due to TiO in the atmosphere. However,
since it is double-gray, the GCM is only using the absorption coefficients to recreate the radiative state of the
atmosphere, and is generally agnostic to the atmospheric composition.}

Since WASP-76b is too hot for \texttt{rate}, we use GGchem to compute
molecular abundances. We then use TauREx to calculate grid emission,
including opacity from H$_2$O, CO, CO$_2$, CH$_4$, HCN, NH$_3$,
C$_2$H$_2$, and C$_2$H$_4$, and integrate the planet emission following the
visibility function described in Equation \ref{eqn:vis}. The radiative
transfer includes no opacity from clouds, because ultra-hot Jupiters
are not expected to form clouds, especially on the dayside
\citep{HellingEtal2021aapCloudTrends, RomanEtal2021apj3DClouds}. 
This is the same process used
in the light-curve modeling, except the GCM has a higher grid
resolution and the temperatures are set by a circulation model instead
of the vertical placement of thermal maps. \change{Our forward model (GCM and 
radiative transfer) and retrieval do not explicitly include TiO opacity, so
our results are still self-consistent with the ground truth.}

For the light-curve simulation, we assume the system parameters listed
in Table \ref{tbl:w76}. We choose a wavelength range of 1.0 -- 2.5
\microns, roughly equivalent to the first order of the JWST NIRISS
single-object slitless spectroscopy observing mode, \change{a} recommended
instrument and mode for transiting exoplanet observations. Using the
JWST Exposure Time Calculator\footnote{https://jwst.etc.stsci.edu}, we
determine the optimal observing strategy is 5 groups per integration,
for a total exposure time of 13.30 s. We assume an observation from
0.4 to 0.6 orbital phase with this exposure time for a total of 2352
exposures over 8.69 hours.

\begin{deluxetable}{cc}
  \tablecaption{WASP-76 System Parameters  \label{tbl:w76}}
  \tablehead{
    \colhead{Parameter} & \colhead{Value}
  }
    \startdata
    Stellar mass, $M_s$        & 1.46 $M_{\odot}$ \\
    Stellar radius, $R_s$      & 1.73 $R_{\odot}$ \\
    Stellar temperature, $T_s$ & 6250 K \\
    Planetary mass, $M_p$      & 0.92 $M_{\textrm{J}}$ \\
    Planetary radius, $R_p$    & 1.83 $R_{\textrm{J}}$\tablenotemark{a} \\
    Orbital period, $P$        & 1.809866 days \\
    Eccentricity, $e$          & 0.0 \\
    Inclination, $i$           & 88.0$^\circ$ \\
    Distance, $D$              & 195.3 pc 
    \enddata
    \tablenotetext{a}{We use Jupiter's volumetric mean radius $R_{\textrm{J}} = 6.9911 \times 10^7$ m.}
\end{deluxetable}

We calculate light-curve uncertainties as photon noise for a single
eclipse, assuming a planetary equilibrium temperature of 2190 K. We
divide the spectrum into 5 spectral bins of equal size in wavelength
space and calculate the photon noise of each bin, assuming Planck functions
for planetary and stellar emission. Star-normalized uncertainties range from 26
ppm at 1.14 \microns\ to 55 ppm at 2.35 \microns. These are
optimistic, uncorrelated uncertainties, but until JWST is operational
it is difficult to predict its behavior.

\subsection{2D Maps}

We fit 2D maps to the synthetic light curves using the methods
described in Section \ref{sec:methods}. Through a BIC comparison 
\change{between all ($l_{max}, N$) pairs for each wavelength}, we
determine the best $l_{max}$ and $N$ for each light curve (see Table
\ref{tbl:2dbics}). As expected, goodness-of-fit improves as $N$
increases, and reaches a limit where increasing $l_{max}$ no longer
improves the fit in a substantial way, once the spherical harmonics
capture the observable temperature structure complexity. It is
important to fit each light curve with its own $l_{max}$ and $N$; in
our case, using a single combination of $l_{max}$ and $N$ results in
worse fits (larger BICs) for all five light curves, with a total BIC
$\approx$15 higher than the total BIC achieved with individual
combinations of $l_{max}$ and $N$.

\begin{figure*}
  \includegraphics[width=7in]{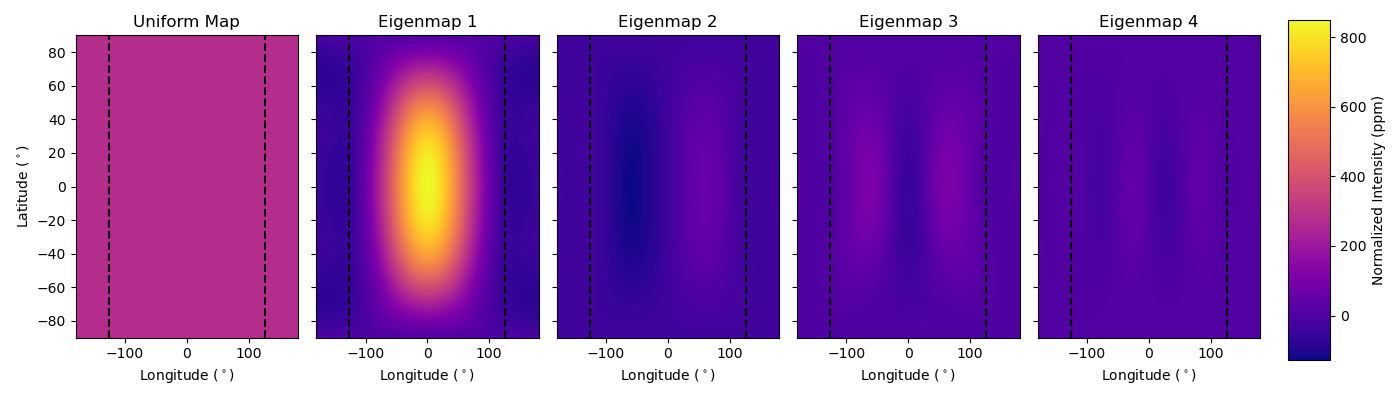}
  \caption{\label{fig:emaps} Uniform map and eigenmap components of
    the 2D fit to the 2.05 \microns\ light curve. The magnitude of
    each map has been scaled by the best-fitting $c_i$ value. The
    vertical dashed lines denote the minimum and maximum visible
    longitudes during the observation, at -126$^\circ$ and
    126$^\circ$.}
\end{figure*}

\begin{figure}
  \includegraphics[width=3.4in]{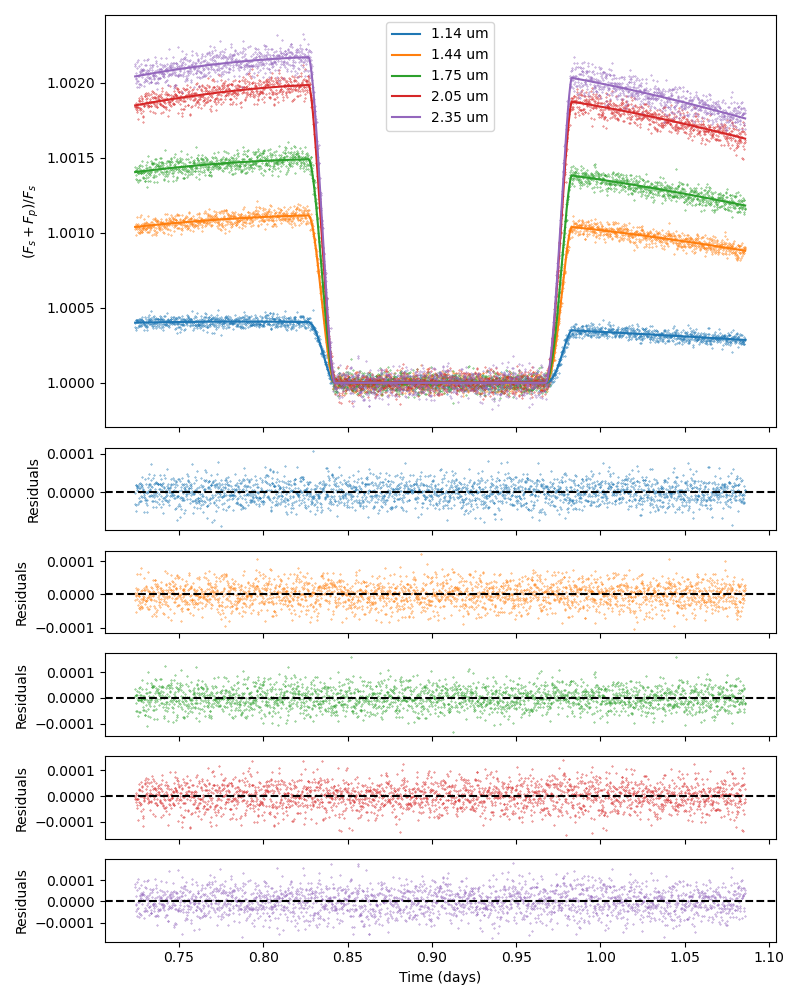}
  \caption{\label{fig:2dlcs} 2D fits to each of the spectroscopic
    light curves, and the residuals.}
\end{figure}

\begin{figure}
  \includegraphics[width=3.4in]{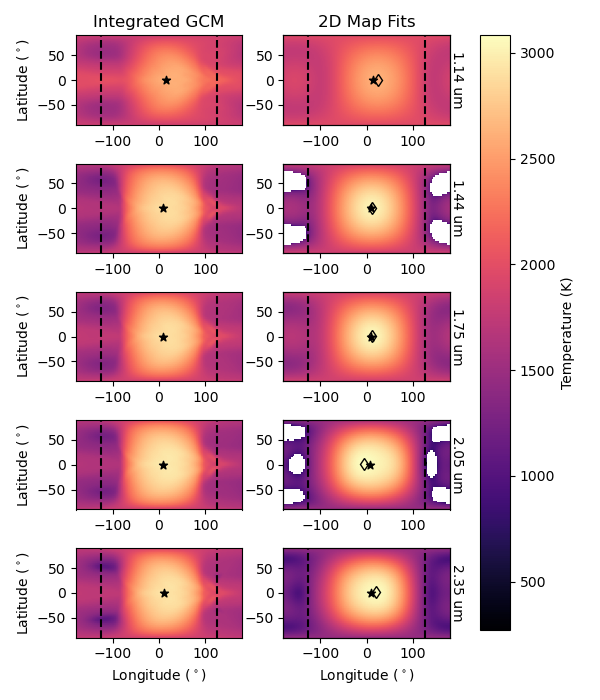}
  \caption{\label{fig:2dmaps} Comparison between input thermal maps
    and retrieved 2D maps. Dashed lines indicate the edges of the
    visible portion of the planet. The GCM hotspot location (see main
    text) is marked with a star and the fitted hotspot location is
    marked with a diamond. \textbf{Left:} Thermal maps from the GCM,
    computed using filter-integrated flux maps and Equation
    \ref{eqn:fmaptotmap}. \textbf{Right:} 2D thermal maps derived from
    fits to each light curve. The maps are plotted at the resolution
    of the GCM, but these maps can be computed at an arbitrary
    resolution. The blank spaces have undefined temperatures, as the
    2D fit produces negative flux in those regions. This is permitted
    because those grid cells are never visible during the
    observation, so we formally have no information about them.}
\end{figure}

The four eigenmaps from the fit to the 2.05 \microns\ data, scaled by
their best-fitting $c_i$ parameters, as well as the scaled uniform map
component, are shown in Figure \ref{fig:emaps}. The first eigencurve
fits the large-scale substellar-terminator contrast, which is extremely
significant for ultra-hot Jupiters and contributes most of the
light-curve flux variation, evident in the significant magnitude of
the scaled eigenmap. The second eigencurve adjusts for the eastward
shift of the planet's hotspot. The third and fourth curves make
further corrections to the temperature variation between the
substellar point and the dayside terminators. Eigenmap information
content decreases with distance from the substellar point, reaching
nearly zero in non-visible portions of the planet, where any variation
is only due to the continuity and integration normalization of the
original spherical harmonics. Figure \ref{fig:2dlcs} shows the light
curves at each wavelength, the best-fitting models, and the
residuals. Figure \ref{fig:2dmaps} shows the best-fitting thermal maps
compared with filter-integrated maps of the GCM.

\begin{deluxetable*}{llrrrrrrrrrr}
  \tablecolumns{12}
  \tablecaption{2D Fit BICs \label{tbl:2dbics}}
  \tablehead{
    \colhead{$l_{max}$} & \colhead{$N$}  & \multicolumn{2}{c}{\change{1.14} \microns} & \multicolumn{2}{c}{1.44 \microns} & \multicolumn{2}{c}{1.75 \microns} & \multicolumn{2}{c}{2.05 \microns} & \multicolumn{2}{c}{2.35 \microns}\\
                       &                & \colhead{Red. $\chi^2$} & \colhead{BIC} & \colhead{Red. $\chi^2$} & \colhead{BIC} & \colhead{Red. $\chi^2$} & \colhead{BIC} & \colhead{Red. $\chi^2$} & \colhead{BIC} & \colhead{Red. $\chi^2$} & \colhead{BIC} 
  }
  \startdata
1 & 1 & {\color{color4} 2.840} & {\color{color4} 6693.59} & {\color{color4} 3.404} & {\color{color4} 8019.42} & {\color{color4} 4.107} & {\color{color4} 9670.25} & {\color{color4} 3.194} & {\color{color4} 7525.70} & {\color{color4} 3.503} & {\color{color4} 8251.41}\\
1 & 2 & {\color{color4} 0.995} & {\color{color4} 2367.63} & {\color{color4} 1.138} & {\color{color4} 2702.60} & {\color{color4} 1.077} & {\color{color4} 2558.76} & {\color{color4} 1.031} & {\color{color4} 2451.88} & {\color{color4} 1.045} & {\color{color4} 2484.05}\\
1 & 3 & {\color{color4} 0.996} & {\color{color4} 2375.45} & {\color{color4} 1.138} & {\color{color4} 2710.44} & {\color{color4} 1.077} & {\color{color4} 2566.49} & {\color{color4} 1.031} & {\color{color4} 2459.67} & {\color{color4} 1.045} & {\color{color4} 2491.97}\\
2 & 1 & {\color{color4} 2.837} & {\color{color4} 6688.25} & {\color{color4} 3.344} & {\color{color4} 7878.70} & {\color{color4} 4.074} & {\color{color4} 9592.54} & {\color{color4} 3.174} & {\color{color4} 7480.07} & {\color{color4} 3.497} & {\color{color4} 8237.88}\\
2 & 2 & {\color{color1} \bf 0.977} & {\color{color1} \bf 2325.42} & {\color{color2} 1.012} & {\color{color2} 2406.48} & {\color{color1} \bf 0.986} & {\color{color1} \bf 2346.08} & {\color{color4} 0.972} & {\color{color4} 2312.66} & {\color{color4} 1.008} & {\color{color4} 2398.75}\\
2 & 3 & {\color{color4} 0.977} & {\color{color4} 2331.34} & {\color{color4} 1.011} & {\color{color4} 2412.64} & {\color{color4} 0.986} & {\color{color4} 2352.35} & {\color{color4} 0.965} & {\color{color4} 2303.82} & {\color{color4} 1.004} & {\color{color4} 2394.72}\\
2 & 4 & {\color{color4} 0.977} & {\color{color4} 2338.40} & {\color{color4} 1.012} & {\color{color4} 2419.98} & {\color{color4} 0.985} & {\color{color4} 2358.07} & {\color{color4} 0.963} & {\color{color4} 2305.38} & {\color{color4} 1.002} & {\color{color4} 2397.02}\\
2 & 5 & {\color{color4} 0.977} & {\color{color4} 2344.87} & {\color{color4} 1.011} & {\color{color4} 2425.29} & {\color{color4} 0.984} & {\color{color4} 2362.29} & {\color{color4} 0.958} & {\color{color4} 2299.91} & {\color{color4} 1.000} & {\color{color4} 2398.90}\\
3 & 1 & {\color{color4} 2.837} & {\color{color4} 6688.27} & {\color{color4} 3.344} & {\color{color4} 7878.49} & {\color{color4} 4.074} & {\color{color4} 9592.37} & {\color{color4} 3.174} & {\color{color4} 7479.63} & {\color{color4} 3.497} & {\color{color4} 8237.61}\\
3 & 2 & {\color{color2} 0.977} & {\color{color2} 2325.88} & {\color{color1} \bf 1.012} & {\color{color1} \bf 2406.48} & {\color{color2} 0.986} & {\color{color2} 2346.64} & {\color{color4} 0.972} & {\color{color4} 2313.66} & {\color{color4} 1.009} & {\color{color4} 2399.74}\\
3 & 3 & {\color{color4} 0.977} & {\color{color4} 2332.19} & {\color{color3} 1.010} & {\color{color3} 2409.55} & {\color{color4} 0.985} & {\color{color4} 2351.54} & {\color{color4} 0.960} & {\color{color4} 2292.15} & {\color{color3} 1.002} & {\color{color3} 2390.80}\\
3 & 4 & {\color{color4} 0.977} & {\color{color4} 2338.15} & {\color{color4} 1.011} & {\color{color4} 2417.44} & {\color{color4} 0.985} & {\color{color4} 2356.39} & {\color{color2} 0.955} & {\color{color2} 2286.45} & {\color{color2} 0.998} & {\color{color2} 2388.35}\\
3 & 5 & {\color{color4} 0.977} & {\color{color4} 2345.42} & {\color{color4} 1.008} & {\color{color4} 2417.71} & {\color{color4} 0.984} & {\color{color4} 2360.85} & {\color{color3} 0.953} & {\color{color3} 2288.38} & {\color{color4} 0.998} & {\color{color4} 2393.68}\\
4 & 1 & {\color{color4} 2.838} & {\color{color4} 6688.84} & {\color{color4} 3.346} & {\color{color4} 7882.25} & {\color{color4} 4.074} & {\color{color4} 9593.78} & {\color{color4} 3.176} & {\color{color4} 7483.10} & {\color{color4} 3.498} & {\color{color4} 8240.50}\\
4 & 2 & {\color{color2} 0.977} & {\color{color2} 2325.51} & {\color{color3} 1.013} & {\color{color3} 2409.68} & {\color{color2} 0.987} & {\color{color2} 2347.56} & {\color{color4} 0.972} & {\color{color4} 2314.35} & {\color{color4} 1.009} & {\color{color4} 2400.42}\\
4 & 3 & {\color{color4} 0.977} & {\color{color4} 2331.51} & {\color{color2} 1.010} & {\color{color2} 2408.14} & {\color{color3} 0.985} & {\color{color3} 2350.48} & {\color{color2} 0.958} & {\color{color2} 2286.61} & {\color{color2} 1.001} & {\color{color2} 2387.66}\\
4 & 4 & {\color{color4} 0.977} & {\color{color4} 2338.46} & {\color{color4} 1.010} & {\color{color4} 2415.93} & {\color{color4} 0.985} & {\color{color4} 2357.49} & {\color{color1} \bf 0.954} & {\color{color1} \bf 2284.98} & {\color{color2} 0.999} & {\color{color2} 2389.33}\\
4 & 5 & {\color{color4} 0.977} & {\color{color4} 2346.11} & {\color{color4} 1.005} & {\color{color4} 2412.12} & {\color{color4} 0.984} & {\color{color4} 2362.43} & {\color{color4} 0.954} & {\color{color4} 2291.29} & {\color{color4} 0.998} & {\color{color4} 2395.44}\\
5 & 1 & {\color{color4} 2.838} & {\color{color4} 6688.86} & {\color{color4} 3.346} & {\color{color4} 7882.38} & {\color{color4} 4.074} & {\color{color4} 9593.83} & {\color{color4} 3.176} & {\color{color4} 7483.09} & {\color{color4} 3.498} & {\color{color4} 8240.57}\\
5 & 2 & {\color{color2} 0.977} & {\color{color2} 2325.53} & {\color{color3} 1.013} & {\color{color3} 2409.79} & {\color{color2} 0.987} & {\color{color2} 2347.61} & {\color{color4} 0.972} & {\color{color4} 2314.36} & {\color{color4} 1.009} & {\color{color4} 2400.49}\\
5 & 3 & {\color{color4} 0.977} & {\color{color4} 2331.53} & {\color{color2} 1.009} & {\color{color2} 2407.95} & {\color{color3} 0.985} & {\color{color3} 2350.36} & {\color{color3} 0.958} & {\color{color3} 2287.06} & {\color{color1} \bf 1.001} & {\color{color1} \bf 2387.66}\\
5 & 4 & {\color{color4} 0.977} & {\color{color4} 2338.33} & {\color{color4} 1.010} & {\color{color4} 2415.73} & {\color{color4} 0.985} & {\color{color4} 2357.50} & {\color{color2} 0.954} & {\color{color2} 2285.61} & {\color{color3} 0.999} & {\color{color3} 2389.76}\\
5 & 5 & {\color{color4} 0.977} & {\color{color4} 2345.92} & {\color{color4} 1.005} & {\color{color4} 2412.08} & {\color{color4} 0.985} & {\color{color4} 2363.12} & {\color{color4} 0.955} & {\color{color4} 2292.65} & {\color{color4} 0.999} & {\color{color4} 2396.47}\\
\enddata
\tablecomments{The optimal fit for each light curve is in black, fits
  with a $\Delta$BIC $\leq 2$ (model preference $\approx$3:1 or better) are in
  blue, fits with $2 < \Delta$BIC $\leq 5$ (model preference $\approx$3:1
  -- 10:1) are in orange, and all worse fits are in red. \change{In some cases
  (e.g., $l_{max} = 2, N = 2$ and $l_{max} = 3, N = 2$ for \change{1.44} \microns), the BICs
  appear to be identical when printed with these digits, but the true best case is in black.}}
\end{deluxetable*}

\subsection{3D Map}

As described above, we fit a 3D model to the spectroscopic light
curves by placing 2D maps vertically in the atmosphere, interpolating
a 3D temperature grid, computing thermochemical equilibrium of the 3D
grid, calculating radiative transfer for each planetary grid cell, and
integrating over the planet for the system geometry of each exposure
in the observation.

Since WASP-76b is an ultra-hot Jupiter (zero-albedo, instantaneous
redistribution equilibrium temperature of 2190 K), we use GGchem to
calculate thermochemical equilibrium abundances. As with the
light-curve generation, the radiative transfer includes H$_2$O, CO,
CO$_2$, CH$_4$, HCN, NH$_3$, C$_2$H$_2$, and C$_2$H$_4$ opacities, all
from ExoTransmit. The atmosphere has 100 discrete layers, evenly
spaced in logarithmic pressure space, from $10^{-6}$ -- $10^2$ bar.

\subsubsection{Spectral Binning}

With spectroscopic observations, the spectral binning is up to the
observer. This choice is far from simple. Smaller bins lead to a
better-sampled spectrum, and each wavelength bin likely probes a
smaller range of pressures, potentially allowing for better
characterization of the atmosphere, as the atmosphere is less likely
to be well fit with all maps placed at similar pressures. However,
small wavelength bins lead to more uncertain light curves and, thus,
more uncertain 2D maps to use in the 3D retrieval. Since we do not
incorporate the uncertainty of the 2D maps in our 3D retrieval (aside
from adjusting their positions vertically), we take the cautious
approach of minimizing this uncertainty by using only five
evenly-spaced spectral bins from 1.0 -- 2.5 \microns.

\subsubsection{Planet Grid Size}
\label{sec:gridsize}

\begin{figure}
  \includegraphics[width=3.4in]{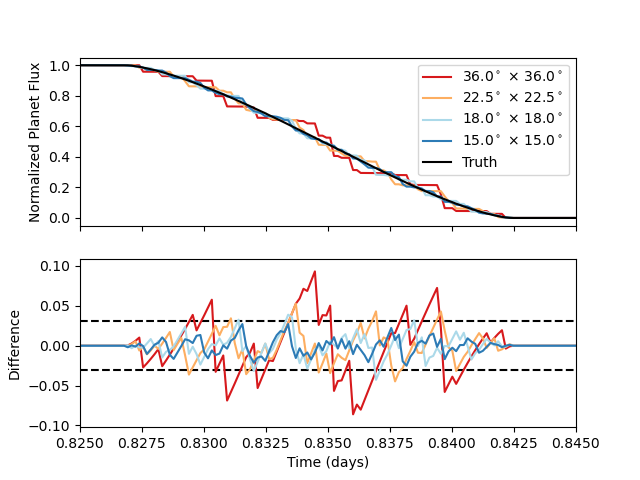}
  \caption{\textbf{Top:} The normalized ingress of a uniformly-bright
    WASP-76b, computed by integrating the planet with the visibility
    function (Equation \ref{eqn:vis}) for a range of grid sizes,
    compared to the true ingress shape computed analytically with
    \texttt{starry}. \textbf{Bottom:} The difference between the true
    ingress and the gridded planets. The dashed lines indicate the
    normalized 1$\sigma$ region for the highest signal-to-noise filter
    in our synthetic observation of WASP-76b. The light curve of a
    sufficiently high-resolution grid should fall within these
    boundaries. \label{fig:gridsize}}
\end{figure}

Our formulation of the visibility function (Equation \ref{eqn:vis}),
counts planet grid cells as either fully visible or obscured depending
on the position of the center of the grid cell relative to the center
of the projected stellar disk. This approximation approaches a smooth
eclipse for infinitely small grid cells, but 3D model calculation time
scales approximately proportionally to the number of grid cells,
quickly becoming infeasible. Therefore we must choose a planetary grid
resolution that adequately captures the eclipse shape within
observational uncertainties and keeps model runtime manageable.
Figure \ref{fig:gridsize} compares uniformly-bright WASP-76b eclipse
ingress models for a range of grid resolutions with the analytic
ingress model. At a resolution of $15^\circ \times 15^\circ$, the
deviations from the analytic model are within the 1$\sigma$ region for
the filter which yields the highest signal-to-noise ratio, so we adopt
this grid resolution.

\subsubsection{Temperature-Pressure Interpolation}

By default, ThERESA uses linear interpolation, in log-pressure space,
to fill in the temperature profiles between the 2D maps. Since the
atmosphere is discretized into pressure layers, if $>2$ thermal maps
are placed between two adjacent pressure layers for a given grid cell,
the grid cells of the central maps of this grouping have no effect on
the 3D thermal structure. This problem is significantly exacerbated
when using the isobaric pressure mapping method, where entire 2D maps,
rather than just a single cell of a map, can be hidden.

We experimented with quadratic and cubic interpolation to mitigate
this problem, as those methods make use of more than just the two
adjacent thermal maps nearest to the interpolation point. However,
these interpolation methods cause significant unrealistic variation in
the vertical temperature structure, often creating negative
temperatures, artificially limiting parameter space (as these negative
temperature models are rejected). Therefore, we elect to use linear
interpolation and rely on the contribution function constraint (see
Section \ref{sec:cffit}) to guide the 2D thermal maps to consistent
pressure levels. While it is still possible to hide maps, for a good
fit this will only happen when maps have similar temperatures.

\subsubsection{Pressure Mapping Function}

We tested each of the pressure mapping functions described in Section
\ref{sec:method3d} -- isobaric, sinusoidal, quadratic, and flexible --
to determine which is most appropriate for our synthetic WASP-76b
observation. For all four cases, we use an additional parameter to
set the planet's internal temperature (at 100 bar) but leave the
upper atmosphere to be isothermal at pressures lower than the highest
2D map in each grid cell.

\begin{figure*}
  \includegraphics[width=7in]{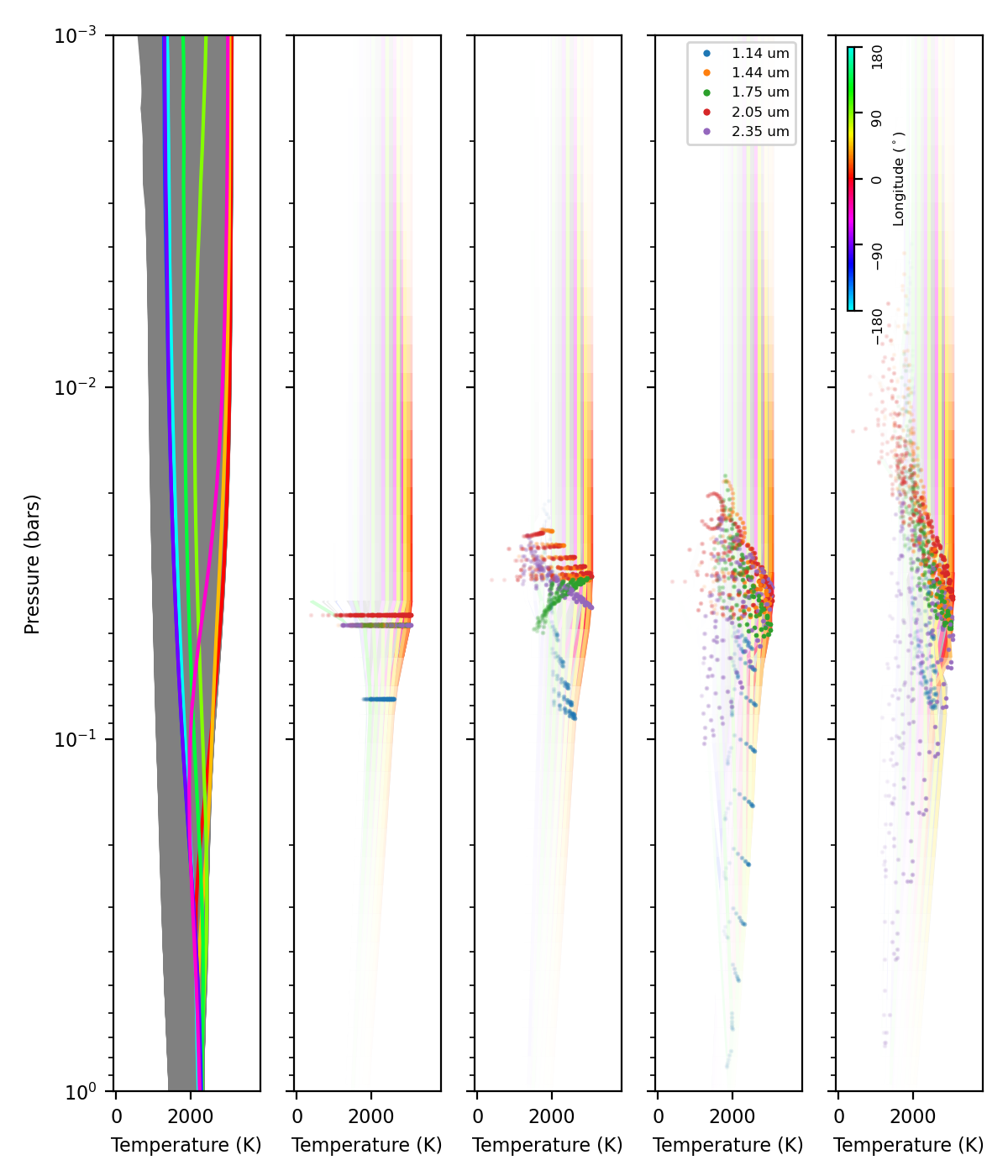}
  \caption{\label{fig:tgrid-comparison} \textbf{Left to right:}
    the input temperature grid from the GCM, the isobaric retrieval,
    the fixed-phase sinusoidal retrieval, the free-phase sinusoidal
    retrieval, and the quadratic retrieval. The
    colored lines indicate temperature profiles along the equator,
    with the GCM downsampled to the spatial resolution of the fitted
    temperature grid. All others are plotted in gray. For the fitted grid,
    plotted color opacity for each profile is weighted by the maximum 
    contribution function (over
    all filters) for each grid cell and pressure layer, normalized to
    the largest contribution. Non-visible grid cells have a
    contribution of 0. Dots indicate the placement of the 2D
    temperature maps for each (visible) grid cell, and their plotting opacity
    is the ratio of the total contribution from that grid cell and
    filter to the maximum total contribution.}
\end{figure*}

The isobaric model, which contains only 6 free parameters -- a
pressure level for each 2D temperature map and an internal temperature
-- fits the data quite well. We achieve a reduced $\chi^2$ of 1.067
and a BIC of 13750.55, including the penalty for contribution function
fitting.  The fit places the 2D maps for the four longest wavelengths,
which all have similar temperatures, at $\approx 0.05$ bar and puts
the \change{1.14} \microns\ map slightly deeper, at $\approx 0.08$ bar,
creating the temperature inversion seen in the GCM (see Figure
\ref{fig:gcm}). Figure \ref{fig:tgrid-comparison} shows the 3D
temperature structure, with each profile weighted by the contribution
functions of each filter in each grid cell.

The sinusoidal pressure mapping function will always result in a fit
at least as good as the isobaric model, as there is a set of
sinusoidal model parameters that replicates the isobaric model, but
the additional complexity may not be justified by the quality of our
data. Here, however, we find a significant improvement. First, we use
the sinusoidal function but fix the phase of the longitudinal sinusoid
($a_4$ in Equation \ref{eqn:sinusoidal}) equal to the hotspot
longitude, effectively assuming the contribution function variation
follows temperature variation. With this parameterization, we find a
reduced $\chi^2$ of 1.013 and a BIC of 13142.84.

If we also allow the $a_4$ terms to vary, the sinusoidal model is no
longer forced to tie the hottest part of each 2D map to the highest or
lowest pressure of that map's vertical position. This added
flexibility improves the fit to a reduced $\chi^2$ of 0.989 and a BIC
of 12879.60. Looking at Figure \ref{fig:tgrid-comparison}, we see that
the hottest portion of 1.75 \microns\ 2D map is shifted to higher
pressures and the \change{1.14} \microns\ map is pushed deeper, smoothing out
the temperature inversion.

The quadratic model is the next step up in complexity. With this
model, we achieve a reduced $\chi^2$ of 1.037 and a BIC of 13580.20.
This is a worse fit than the sinusoidal models, and a much higher BIC,
suggesting that the sinusoidal model better captures the 3D placement
of the 2D maps using fewer free parameters.

We also tested the ``flexible'' 3D model. With 12 latitudes and 24
longitudes, and a visible range of $\phi \in (-126^\circ, 126^\circ)$,
there are 216 visible and partially-visible grid cells. With 5
wavelength bins and an internal temperature parameter, there are 1081
model parameters in total (including the internal temperature). In a
best-case scenario, we would achieve a reduced $\chi^2$ of 1, implying
a BIC $\geq 21882.26$, far greater than the BICs we achieve with
simpler, less flexible models. Without a drastic reduction in
observational uncertainties, the data are unable to support such a
model. In fact, such a reduction in light-curve uncertainties would
require a finer planetary grid (see Section \ref{sec:gridsize}), in
turn requiring additional parameters, further increasing the BIC and
decreasing preference for this model.

\change{As with choosing $N$ in the 2D mapping, we use the BIC to
determine the optimal 3D model. For these data, we achieve the
lowest BIC using the free-phase sinusoidal model.}

\subsection{Credible Region Errors}

To assess the completeness of the MCMC parameter-space exploration,
we compute the Steps Per Effectively Independent Sample (SPEIS),
Effective Sample Size (ESS), and the absolute error on our 68.3\% (1$\sigma$)
credible region $\sigma_{C}$ of our posterior distribution following
\cite{HarringtonEtal2021arxivBART1}. We compute SPEIS using the 
initial positive sequence estimator \citep{Geyer1992statsciMCMC}, then divide 
the total number of iterations by the SPEIS to calculate ESS. Then, 

\begin{equation}
    \sigma_{C} = \sqrt{\frac{C (1 - C)}{{\rm ESS} + 3}},
    \label{eqn:cr}
\end{equation}

\noindent
where $C = 0.683$ is a given credible region. We calculate an 
ESS for each parameter of each chain and sum over all chains to 
get a total ESS.

\section{Discussion}
\label{sec:disc}

\subsection{2D Thermal Maps}

From a qualitative perspective the retrieved maps match the GCM well.
Figure \ref{fig:2dmaps} shows a comparison of the 2D thermal maps
extracted from the GCM used to generate the light curves and the
retrieved maps. Large-scale features, like the planet's hotspot and
its eastward shift from the substellar point are recovered. As
expected, there is little match between the night side of the GCM and
the fitted maps, and the fits are often non-physical in this region,
as we do not apply physicality constraints to non-visible portions of
the planet. The fits are also unable to recover smaller-scale
features, like the latitudinal width of the equatorial jet
and the sharp drops in temperature near the planet's terminator, which
cannot be replicated without many additional eigencurves.

\begin{figure}
  \includegraphics[width=3.4in]{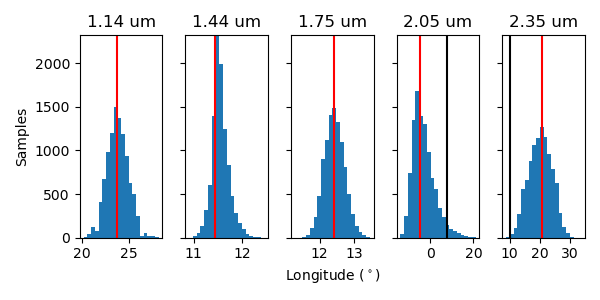}
  \caption{\label{fig:hshist} Histograms of hotspot longitude from the
    MCMC posterior distributions for each 2D map. The best-fitting
    hostpot longitude is overplotted in a red line. The GCM hotspot
    longitude is shown with a black line if it falls within the
    longitudes of the $x$-axes.}
\end{figure}

From the MCMC parameter posterior distribution, we can generate a
posterior distribution of maps to examine a variety of features. By
finding the maximum of a subsample of 10,000 maps in the posterior
distribution, we can identify the location and uncertainty of the
planet's hotspot and determine if there is a longitudinal shift from
the substellar point, a useful metric for understanding the dynamics
of the planet's atmosphere. For WASP-76b, we determine hotspot
longitudes of $23.8^{+1.1}_{-1.2}$, $11.4^{+0.3}_{-0.1}$,
$12.4^{+0.3}_{-0.3}$, $-4.9^{+4.8}_{-4.2}$, and $20.8^{+3.6}_{-4.5}$
degrees for each bandpass in increasing wavelength, respectively (see
Figure \ref{fig:hshist}). For comparison to the GCM, we constructed 1D
cubic splines along the equators of the GCM temperature maps (Figure
\ref{fig:2dmaps}) and found their maxima. The GCM has hotspot
longitudes of $14.1^\circ$, $9.1^\circ$, $9.2^\circ$, $7.8^\circ$, and
$10.1^\circ$. In a broad sense, our 2D retrieval captures the stronger
eastward shift at \change{1.14} \microns\ and places the other 4 maps closer
together, much like the GCM maps. Differences are likely due to the
limited temperature structures possible with $N \leq 4$ fits and
approximating the GCM hotspot as the equatorial longitude with the
peak temperature. Unsurprisingly, the fitted hotspot latitudes are
very near the equator, since none of the included eigenmaps contain
significant north-south variation, which should be detectable for
orbits with non-zero (but less than $1 - R_p/R_s$) impact parameters.

\begin{figure}
  \includegraphics[width=3.4in]{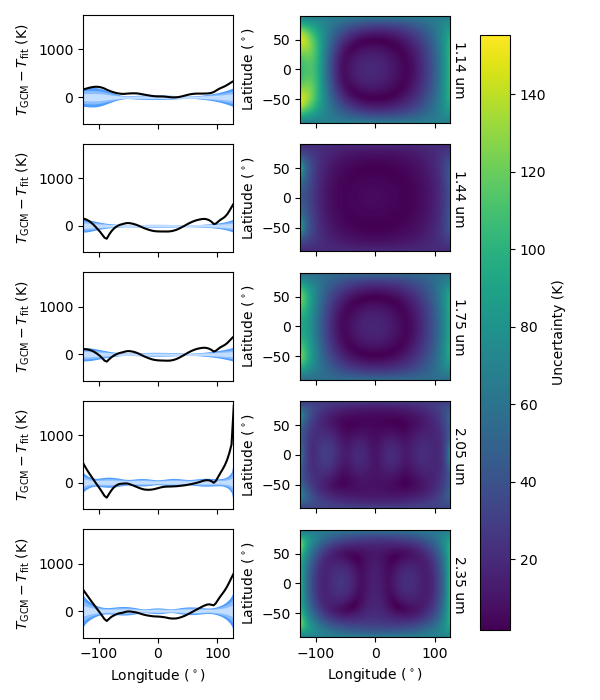}
  \caption{\label{fig:tmapunc} Uncertainties in the 2D temperature map
    fits, calculated as the standard deviation of the posterior
    distribution of maps. Plots are at the spatial resolution of the
    GCM, although they can be computed at any resolution. Axes are
    limited to locations visible during the observation.
    \textbf{Left:} Difference between the GCM (in black) and fitted
    equatorial
    temperatures. The shaded regions denote the 1$\sigma$, 2$\sigma$,
    and 3$\sigma$ boundaries. \textbf{Right:} Temperature uncertainty
    maps, calculated as the standard deviation at each location.}
\end{figure}

Using the subsample of posterior maps, we also can calculate
uncertainties on the temperature maps, at arbitrary resolution, by
evaluating the standard deviation of the temperature in each grid
cell. Figure \ref{fig:tmapunc} shows the full temperature uncertainty
maps and equatorial profiles. Uncertainties are low, at $\approx 10$
K, for grid cells that are visible throughout the observation. Grid
cells at higher longitudes are less well constrained because 1) they
are only visible or partially visible for a portion of the
observation, and 2) they contribute less to the planet-integrated flux
due to the visibility function (Equation \ref{eqn:vis}). Likewise,
high latitudes are less well constrained.

\begin{figure*}
  \includegraphics[width=7in]{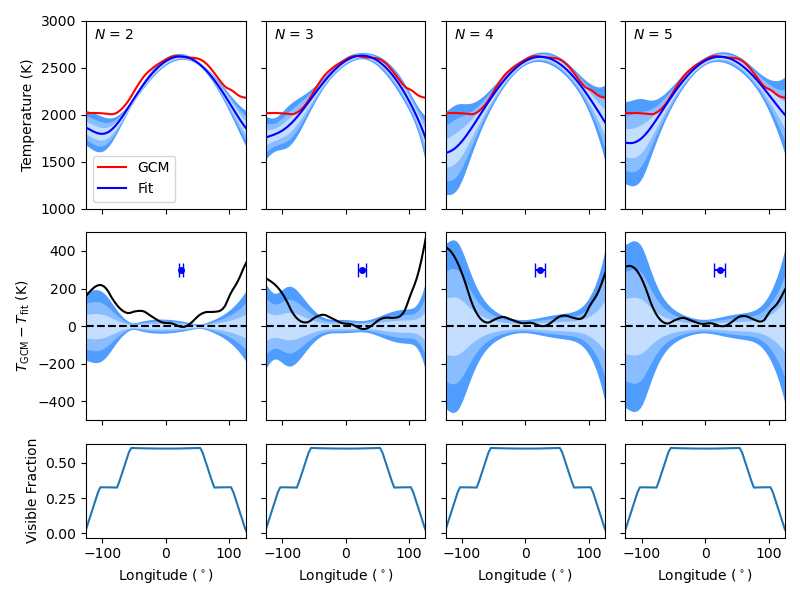}
  \caption{\label{fig:unc-v-n} A comparison of the equatorial
    temperature uncertainties on \change{1.14} \microns\ maps retrieved with
    different numbers of eigencurves. The blue region denotes 1, 2,
    and 3$\sigma$. Only visible longitudes are shown. \textbf{Top:}
    The equatorial temperatures of the GCM and the best
    fit. \textbf{Middle:} The difference between the GCM and fitted
    equatorial temperatures. The blue data point shows the best-fit
    hotspot longitude and its 3$\sigma$ uncertainty ($y$ location
    arbitrary). \textbf{Bottom:} The fraction of the observation
    during which a given longitude, along the equator, is visible to
    any degree, including the effects of both planet rotation and the
    occultation by the star. \change{The plateau centered on $0^\circ$
    shows the longitudes which are visible for the whole observation, except
    during eclipse. The smaller plateaus near $-90^\circ$ and $90^\circ$ are
    the longitudes visible for the entirety of pre- or post-eclipse but not 
    vice versa due to planet rotation during eclipse.}}
\end{figure*}

While the thermal maps agree with the GCM at many locations on the
planet, there are a few places where the discrepancy is significant
within the uncertainties. In several bandpasses we overestimate the
temperature of the hotspot, four of the maps overestimate the
temperature of the western terminator, and all five maps underestimate
temperatures at the extreme ends of the visible longitudes. These
discrepancies can all be explained by the simplicity of the 2D model,
which only contains two to four statistically-justified
eigencurves. The model cannot capture the fine details of the GCM
thermal structure, and the uncertainties are likewise
constrained. Indeed, if we increase the number of eigencurves, we both
improve the match with the GCM and increase the temperature
uncertainties to encompass the difference between the fit and truth
(see Figure \ref{fig:unc-v-n}). However, this also significantly
increases the uncertainty on the location of the planet's hotspot and
the additional parameters are not statistically justified, evident in
the minor changes to the best fit as more parameters are
introduced. For example, adding only one more eigencurve to the \change{1.14}
\microns\ fit ($N = 3$) increases the BIC by $5.92$ (Table
\ref{tbl:2dbics}), a preference for the $N = 2$ model of
$\approx$19:1. This figure also shows the fraction of the observation
where each equatorial longitude is visible, which highlights the
correlation between the observability of a location on the planet and
the uncertainty on the retrieved temperature of that location. When
interpreting analyses of real data we must be aware of these
limitations.

\subsection{3D Retrieval}

At first glance, the best-fitting 3D models all appear to be
physically unrealistic, with internal temperatures at $\approx 500$ K,
much lower than the GCM internal temperature of $\approx 3000$
K. However, if we examine the contribution functions of the
best-fitting model (shown in line opacity in Figure
\ref{fig:tgrid-comparison}), we see that our spectral bins are
primarily sensitive to pressures from 0.001 -- 1 bar, and the
majority of the emitted flux comes from the hottest grid cells at
pressures between 0.01 and 0.1 bar, far from the interior of the
planet at 100 bar. Therefore, the internal temperature parameter is
only affecting the emitted flux by controlling the temperature
gradient near the deepest thermal map, and the absolute value of the
parameter is unimportant. In all our fits, this extends the thermal
inversion to the deepest visible pressure layers, consistent with the
thermal inversion in the GCM, which continues down to $\approx 0.5$
bar at the substellar point. The temperature structure as a whole is
physically implausible at higher pressures, but the portions sensed by
the observation are reasonable and similar to the GCM. \change{We
note that our MCMC analyses started from a more plausible internal
temperature of 3000 K, using a uniform prior over the range [0, 4000] K, 
but without wavelengths which probe the
deep atmosphere, those fits were quickly ruled out in favor of the
fits presented here.}

\begin{figure}
    \centering
    \includegraphics[width=3.25in]{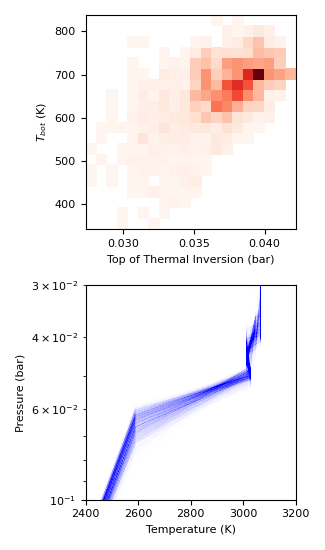}
    \caption{\label{fig:tbot-v-invp} MCMC posterior distribution of temperature
    profiles at the substellar point, showing our parameterization scheme
    is not influencing the location of the thermal inversion. For computational
    feasibility, we thinned the posterior distribution by a factor of 100.
    \textbf{Top:} The minor positive correlation between $T_{bot}$ 
    and the pressure at the top of the thermal inversion, where the
    temperature gradient becomes negative. 
    \textbf{Bottom:} The distribution of temperature-pressure
    profiles in the MCMC posterior near the thermal inversion at the substellar 
    point. At lower pressures the atmosphere is isothermal and at higher pressures
    the linear trend continues down to $T_{bot}$ at 100 bar.
    }
\end{figure}

\change{
To further understand the effects of the $T_{bot}$ parameter, we
examined the thermal profiles in the MCMC posterior as a function
of $T_{bot}$. First, we looked for potential correlations
between $T_{bot}$ and the location of the thermal inversion, measured
by evaluating the 3D thermal profiles at high vertical (pressure) resolution
and finding the pressure level where the temperature gradient becomes
negative. This relationship, for the substellar point, is shown in
Figure \ref{fig:tbot-v-invp}, which shows a minor positive correlation,
although the variation in the inversion pressure is small. Thus,
the $T_{bot}$ parameter is not affecting the location of the
thermal inversion.
}

There are many similarities between the isobaric fit and the two
sinusoidal fits. In all cases, the \change{1.14} \microns\ temperature map is
placed at higher pressures than the others to create the thermal
inversion near \change{0.04} bar and the planet has a low internal temperature
to continue the temperature inversion. The upper atmosphere is roughly
the same, especially in the hottest grid cells. However, while the
isobaric model is forced to place all maps near their peak
contributions at the planet's hotspot, the additional flexibility of
the sinusoidal model allows the temperature maps to better match their
contribution functions (and the emitted spectra) across the entire
planet. This is evident in the significantly improved $\chi^2$ and
BIC. We take the free-phase sinusoidal model to be the optimal fit.

\begin{figure*}
  \includegraphics[width=3.5in]{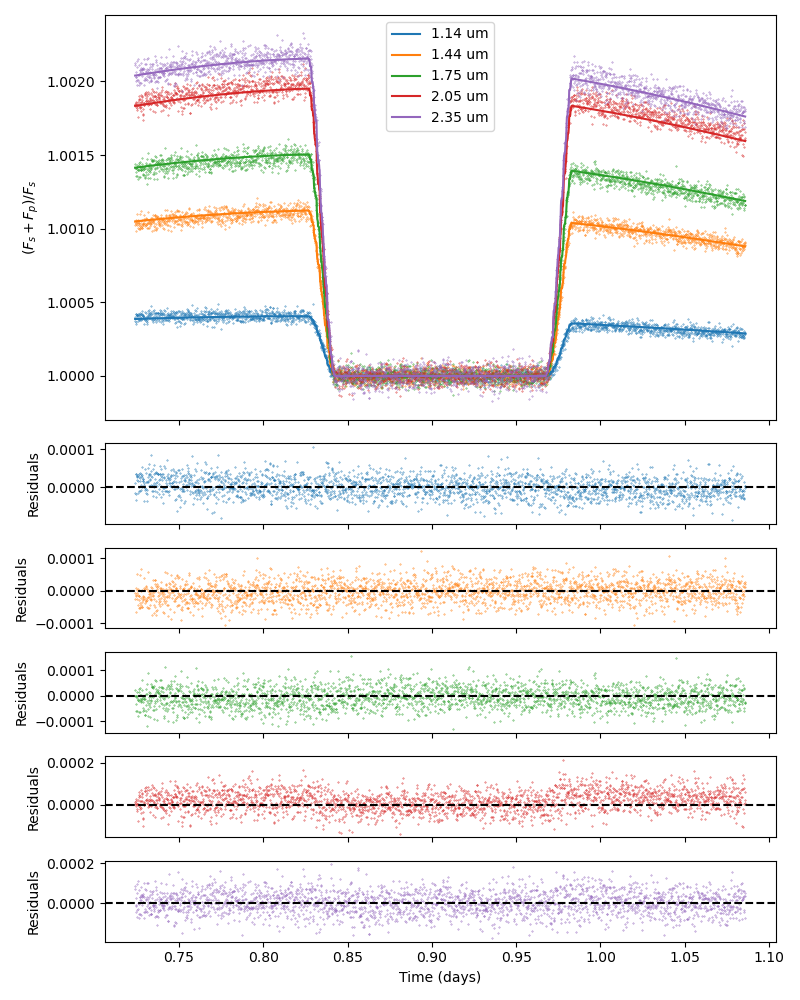}
  \includegraphics[width=3.5in]{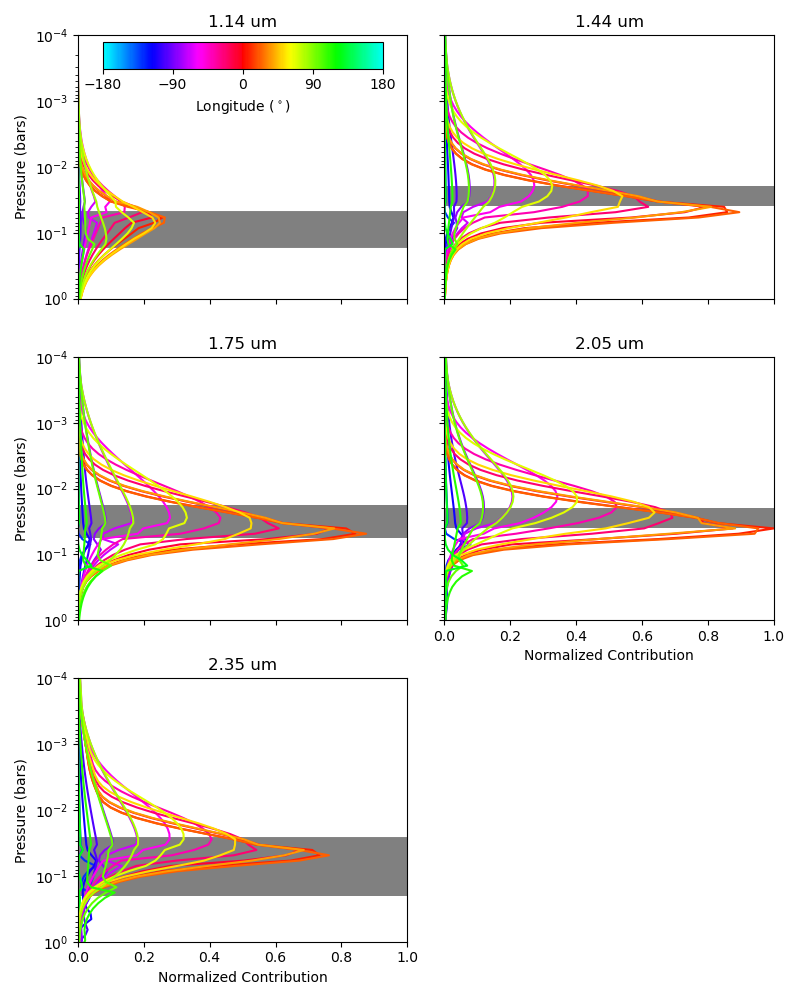}
  \caption{\label{fig:3dlcs} \textbf{Left:} The light-curve fit from
    the best-fitting 3D model, and the residuals. \textbf{Right:}
    Contribution functions, normalized to the maximum contribution,
    for visible grid cells along the equator in the best-fitting 3D
    model. Each panel shows a different filter. The gray boxes
    indicate the range of pressures covered by the respective 2D map.}
\end{figure*}

Figure \ref{fig:3dlcs} shows the optimal 3D model fit to the light
curves, and Table \ref{tbl:3dfit} lists the model parameters
with their 1$\sigma$ credible regions, SPEIS, ESS, and $\sigma_C$, 
from a 28-day run with $\approx810,000$ iterations over 7 Markov
chains. We discard the first 80,000 iterations of each chain, as 
the $\chi^2$ was still improving significantly, so these iterations
are not representative of the true parameter space. \change{These 
parameters give information about the pressures probed by each filter, 
as follows: $a_1$ is the average logarithmic 
pressure probed by the corresponding temperature map; $a_2$ is
the change in logarithmic pressure probed from the equator to the poles; and
$a_3$ is a similar change from the longitude $a_4$ to $a_4$ + 180$^\circ$. 
However, one must keep in mind that the observation is not sensitive
to all longitudes, so the $a_3$ parameter should only be considered
representative of the longitudinal change in probed pressure within
the range of visible longitudes (-126$^\circ$ to 126$^\circ$ for this
observation).}

\begin{deluxetable}{lrrrrr}
  \tablecolumns{6}
  \tablecaption{\change{Free-phase Sinusoidal} 3D Model Parameters \label{tbl:3dfit}}
  \tablehead{
    Name\change{\tablenotemark{a}} & Best fit & $C = 0.683$ & SPEIS\tablenotemark{b} & ESS\tablenotemark{c} & $\sigma_C$
  }
    \startdata
\multicolumn{6}{c}{\change{1.14} \microns}\\
\hline
$a_1$ & -0.68 & [-0.87, -0.61] &  5656 & 143.2 & 3.8\%\\
$a_2$ & 0.05 & [-0.06, 0.17] &  6586 & 123.0 & 4.1\%\\
$a_3$ & 0.65 & [0.45, 0.67] & 11708 &  69.2 & 5.5\%\\
$a_4$ ($^\circ$) & 159 & [153, 161] & 10720 &  75.6 & 5.2\%\\
\hline
\multicolumn{6}{c}{\change{1.44} \microns}\\
\hline
$a_1$ & -1.73 & [-1.78, -1.67] &  5972 & 135.7 & 4.0\%\\
$a_2$ & 0.33 & [0.30, 0.44] &  5907 & 137.2 & 3.9\%\\
$a_3$ & -0.05 & [-0.04, -0.00] & 21794 &  37.2 & 7.3\%\\
$a_4$ ($^\circ$) & 109 & [20, 127] & 23214 &  34.9 & 7.6\%\\
\hline
\multicolumn{6}{c}{1.75 \microns}\\
\hline
$a_1$ & -1.69 & [-1.73, -1.62] &  5389 & 150.3 & 3.8\%\\
$a_2$ & 0.29 & [0.21, 0.35] &  5867 & 138.1 & 3.9\%\\
$a_3$ & 0.11 & [0.09, 0.13] &  7082 & 114.4 & 4.3\%\\
$a_4$ ($^\circ$) & -23 & [-25, -11] & 13547 &  59.8 & 5.9\%\\
\hline
\multicolumn{6}{c}{2.05 \microns}\\
\hline
$a_1$ & -1.69 & [-1.69, -1.57] &  6750 & 120.0 & 4.2\%\\
$a_2$ & 0.28 & [0.15, 0.29] &  6436 & 125.9 & 4.1\%\\
$a_3$ & -0.05 & [-0.05, -0.02] & 11513 &  70.4 & 5.4\%\\
$a_4$ ($^\circ$) & 95 & [65, 98] & 19405 &  41.7 & 7.0\%\\
\hline
\multicolumn{6}{c}{2.35 \microns}\\
\hline
$a_1$ & -1.43 & [-1.51, -1.33] &  8576 &  94.5 & 4.7\%\\
$a_2$ & 0.22 & [0.10, 0.26] &  6472 & 125.2 & 4.1\%\\
$a_3$ & -0.23 & [-0.25, -0.15] & 10128 &  80.0 & 5.1\%\\
$a_4$ ($^\circ$) & 39 & [32, 42] & 15364 &  52.7 & 6.2\%\\
\hline
$T_{{bot}}$ (K) & 701 & [619, 737] &  6137 & 132.0 & 4.0\%\\
    \enddata
\tablenotetext{a}{\change{$a_1$ is the average vertical location of the 
temperature map, $a_2$ is the sinusoidal amplitude of variation in pressures probed 
by the map with latitude, $a_3$ is the sinusoidal amplitude of variation in pressures
probed with longitude, and $a_4$ is the phase shift of the longitudinal sinusoid. 
All quantities are in log(pressure), in bars. See Equation \ref{eqn:sinusoidal}}}
\tablenotetext{b}{\change{Steps Per Effectively Independent Sample}}
\tablenotetext{c}{\change{Effective Sample Size}}
\end{deluxetable}

Since the noise in the observation is purely uncorrelated, we
can easily see where the model fails to match the observation.  There
is a minor negative slope in the residuals at \change{1.14} \microns,
indicating extra flux east of the substellar point and a flux
deficiency west of the substellar point, at those wavelengths. This
agrees with the more eastern hotspots found in our 2D maps. Our best
fit also produces slightly less flux in the 2.05 \microns\ band than
the GCM. All these differences are well within the uncertainties,
however, as indicated by the low $\chi^2$. Increasing $N$ in the 2D
fit could allow for fine adjustment of the 3D thermal structure and
improve the 3D fit in these areas, but, as discussed above, additional
free parameters are not justified by the data uncertainties. \change{Fits
using the simpler pressure-mapping functions have slightly stronger variations
in the residuals, but very similar shapes (e.g., a stronger slope at 1.14 \microns).}

\begin{figure*}
  \includegraphics[width=7in]{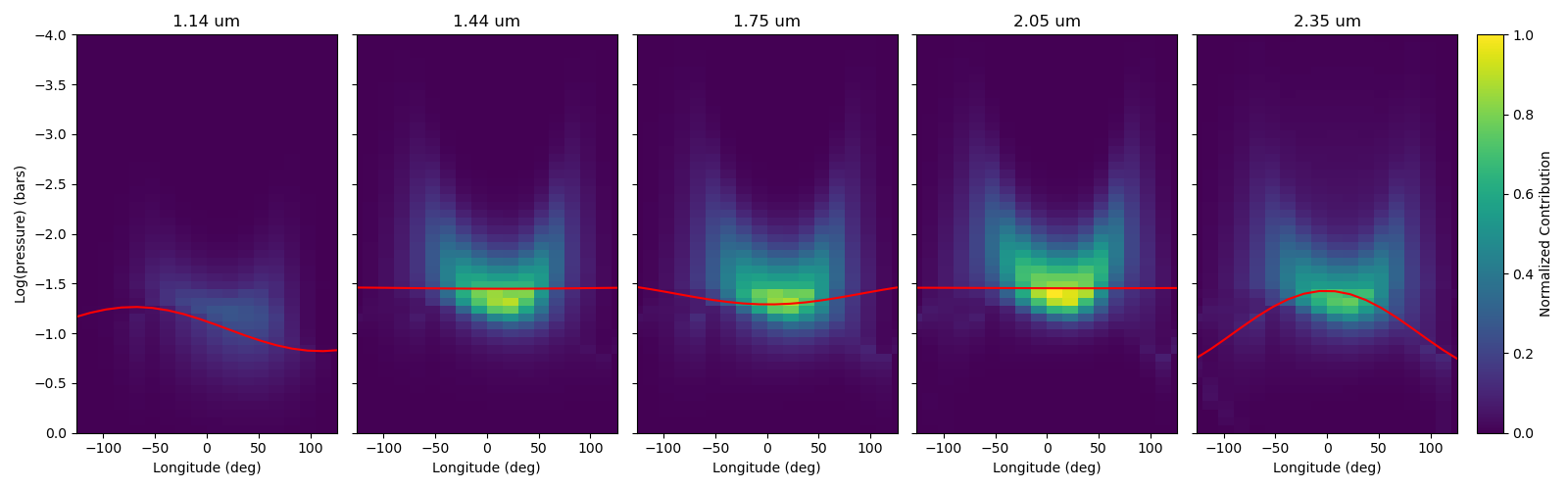}
  \caption{\label{fig:cf-slice} The normalized equatorial contribution
    functions for each spectroscopic bin. The vertical, equatorial
    placements of the 2D thermal maps \change{from the optimal fit
    of the free-phase sinusoidal function (Table \ref{tbl:3dfit})} 
    are overplotted in red.}
\end{figure*}

This figure also shows a comparison between the 3D model contribution
functions and the vertical placement of the 2D thermal maps, to
examine the effectiveness of our contribution function consistency
requirement (Section \ref{sec:cffit}). The range of vertical placement
of the 2D thermal maps (gray boxes) align well with the peaks of the
contribution functions, demonstrating that our $\chi^2$ penalty is
successfully guiding the fit to a consistent result. Figure
\ref{fig:cf-slice} shows a comparison of the equatorial contribution
functions with the 2D map placements overlaid, which confirms that the
2D maps' vertical position matches the contribution functions,
especially for the hottest grid cells.

Uncertainties on the placement of the 2D maps are straightforward to
evaluate, using the marginalized MCMC posterior distributions. For
instance, for the isobaric model, the 2D maps are placed, in order of
increasing wavelength, at $-1.115^{+0.006}_{-0.014}$,
$-1.326^{+0.004}_{-0.017}$, $-1.325^{+0.011}_{-0.001}$,
$-1.353^{+0.010}_{-0.023}$, and $-1.326^{+0.005}_{-0.014}$ log(bar
pressure), with an internal temperature of $630^{+64}_{-38}$ K. \change{The
uncertainties on these model parameters, and those of the other 3D model
functions, can be lower than the model's vertical resolution since the discretized
layers of the model have temperatures interpolated from the vertical placement
of the 2D temperature maps. The 2D maps can be placed anywhere within the 
pressure range of the model, and are not restricted to the precise layering
of the model.}

However, uncertainties on the 3D temperature model as a whole are more
nuanced. A distribution of 3D models generated from the MCMC parameter
posterior distribution will lead one to believe the atmosphere is very
well constrained. However, much like the uncertainties on the 2D
models are restricted by the flexibility of the eigenmaps,
uncertainties on the best-fitting 3D models are restricted by the
models' functional forms. \change{Additionally, the uncertainties
on the 2D maps are not propagated to the 3D fits.} The upper atmosphere 
will appear constrained
to absolute certainty because the model sets these temperatures to be
isothermal with the lowest-pressure 2D map in each grid
cell. Therefore, temperatures at these low pressures do not vary in
the MCMC unless the 2D maps are swapping positions, and even then the
upper atmosphere of each grid cell can only have a number of unique
temperatures equal to the number of 2D maps present. For pressures
between 2D map placements, the temperature profile interpolation and
vertical shifting of the 2D maps leads to a range of temperatures in
the posterior distribution of models, but variation is still
significantly limited by the temperatures present in the 2D maps. At
higher pressures, temperatures are strongly tied to the deepest 2D map
but some variation is allowed by the internal temperature
parameter. Uncertainties on the 3D model are much better understood by
studying the contribution functions, as uncertainties should be
considered unconstrained in portions of the atmosphere with zero
contribution to the planet's flux.

\subsubsection{Thermal Inversion}
\change{
The presence of strong optical absorbers, such as TiO,
is often invoked to explain the presence of thermal inversions,
as this molecule can significantly heat a planet's upper atmosphere
\citep{HubenyEtal2003apjTiOandVO, FortneyEtal2006apjHD149,
FortneyEtal2008apjJupiters}. Evidence of TiO has been found in
observations of multiple hot Jupiters (e.g., \citealp{KirkEtal2021ajWASP103b,
ChangeatEdwards2021apjlKELT9b}), including WASP-76b \citep{FuEtal2021ajWASP76b}.
}

\change{
Both our radiative transfer forward
model of the ground truth and the retrieved 3D model do not include
TiO, and yet we retrieve the input thermal inversion.
As described in Section \ref{sec:w76b}, the temperature inversion in the GCM
is produced by the relative ratio between the infrared and optical 
absorption coefficients, which are broadband and do not include 
molecular spectral absorption (such as by TiO). The retrieval framework
does not require TiO, but for a different reason; the temperature
profiles are entirely parameterized and so do not require a physical
heating mechanism to produce an inversion. This is a strength of the model,
as real exoplanet atmospheres are likely in complex thermal states which
can, in principle, be recreated by our model. 
}

\change{
While we achieve an excellent fit to the eclipse spectra without TiO, 
it is possible that the inclusion of this molecule would affect
the emission spectra enough to shift the pressures probed by
our observations, changing the optimal locations of the 2D temperature
maps. However, TiO opacity peaks in the optical and 
falls off rapidly into the near-infrared wavelengths (e.g.,
\citealp{Gharib-NezhadEtal2021apjsEXOPLINES}). Thus, the impact
on our models should be small.
}
\section{Conclusions}
\label{sec:conc}

In preparation for the 3D exoplanet mapping capabilities of JWST, we
have presented the ThERESA code, a fast, public, open-source package
for 3D exoplanet atmosphere retrieval. The code builds upon the
maximally-informative 2D mapping techniques of
\cite{RauscherEtal2018ajMap} with the addition of 3D planet models,
composition calculations, radiative transfer computation, and planet
integration to model observed eclipse spectra. Thus, we combine a 2D
mapping scheme with 1D radiative transfer into a 3D model with a
manageable number of parameters, enabling fast fit convergence. For
example, the isobaric model, running on 11 processors and $<5$
gigabytes of memory, with 12 latitudes, 24 longitudes, 100 pressure
layers, and eight molecules, converges in $<3$ days of runtime. More
complex models can take a few weeks depending on the desired error
level in the parameter credible regions.

ThERESA improves upon the 2D mapping methods of
\cite{RauscherEtal2018ajMap} by (1) using TSVD PCA to ensure that
eigencurves have the expected zero flux during eclipse, reducing the
need for a stellar correction term, and (2) restricting parameter
space to avoid non-physical negative fluxes at visible locations on
the planet, forcing positive temperatures and enabling radiative
transfer calculations. Through a reanalysis of \textit{Spitzer} HD
189733 b observations, we demonstrated the accuracy of ThERESA's
implementation of 2D eigencurve mapping. Our measurement of the
eastward shift of the planet's hotspot agrees extremely well with
previous studies \citep{MajeauEtal2012apjlHD189Map,
  RauscherEtal2018ajMap}.

Our 3D planet models consist of functions which attach 2D maps to
pressures that can vary by position on the planet, using functions
which range in complexity from a single pressure per map to a
maximally-flexible model with a parameter for the vertical position of
every 2D map in every grid cell. We also require that 2D maps be
placed near the peaks of their corresponding contribution functions,
ensuring consistency between the 3D model and the radiative transfer
calculations.

To test the accuracy of our retrieval method, and to explore the
capabilities of eclipse mapping with JWST, we generated a synthetic
eclipse observation from WASP-76b GCM results. Our 2D maps are able to
retrieve the large-scale thermal structure of the GCM with $l_{\rm
  max} \leq 5$ and $N \leq 4$, the highest-complexity fits justified by a BIC
comparison.  We caution that the limited complexity of the eigencurves
and eigenmaps limits the structures possible in the best-fitting maps
and their uncertainties. Future mapping analyses must be cognizant of
these limitations when presenting results.

Our 3D retrievals, regardless of the temperature-to-pressure model
used, were able to accurately determine the temperatures of the
planet's \change{atmosphere that we are sensitive to, from 
$\sim0.001 - 0.1$ bar, including} the presence of a thermal
inversion at the planet's hotspot near \change{0.04} bar, while maintaining a
radiatively consistent atmosphere \change{by ensuring that 3D model
contribution functions match the vertical placements of the 2D 
temperature maps}. Through a BIC comparison, a
sinusoidal model function that includes combined latitudinal,
longitudinal, and vertical information was preferred over a purely
isobaric model, demonstrating that 3D models are necessary to
interpret JWST-like observations, and ThERESA can perform the analyses
on such data.

\begin{acknowledgments}

  We thank the anonymous referee for their insightful comments which
  led to improvements to the manuscript.
  We thank Eric Agol and Nicolas Cowan for the reduced HD 189733 b data.
  We thank contributors to SciPy, Matplotlib, Numpy, and the Python
  Programming Language. Part of this work is based on observations
  made with the \textit{Spitzer Space Telescope}, which was operated
  by the Jet Propulsion Laboratory, California Institute of
  Technology, under a contract with NASA. This research was supported
  in part through computational resources and services provided by
  Advanced Research Computing at the University of Michigan, Ann
  Arbor. This work was supported by a grant from the Research Corporation 
  for Science Advancement, through their Cottrell Scholar Award.
  
\end{acknowledgments}

\software{NumPy \citep{HarrisEtal2020natNumPy}, Matplotlib
  \citep{Hunter2007cseMatplotlib}, SciPy
  \citep{VirtanenEtal2020natmSciPy}, Scikit-learn,
  \citep{PedregosaEtal2011jmlrScikitLearn}, starry
  \citep{LugerEtal2019ajStarry}, MC3
  \citep{CubillosEtal2017ajRedNoise}, Tau-REx 3
  \citep{Al-RefaieEtal2019arxivTauRExIII}, RATE
  \citep{CubillosEtal2019apjRATE}, GGchem
  \citep{WoitkeEtal2018aandaGGchem}}

\bibliography{theresa.bib}

\end{document}